\documentclass[a4paper,11pt]{article}
\pdfoutput=1 % if your are submitting a pdflatex (i.e. if you have
\usepackage{jcappub} % for details on the use of the package, please
\usepackage[T1]{fontenc} % if needed
\usepackage{multirow}

\title{\boldmath Exploring neutrinos from proton decays catalyzed by GUT monopoles in the Sun}

\author[a]{Hang Hu}
\author[b]{Jie Cheng}
\author[c,1]{Wan-Lei Guo \note{Corresponding author.}}
\author[a,d]{Wei Wang}

\affiliation[a]{School of Physics, Sun Yat-Sen University, Guangzhou, 510275, China}
\affiliation[b]{North China Electric Power University, Beijing, 102206, China}
\affiliation[c]{Institute of High Energy Physics, Chinese Academy of Sciences,
Beijing 100049, China}
\affiliation[d]{Sino-French Institute of Nuclear Engineering and Technology, Sun Yat-Sen University, Zhuhai, 519000,China}

\emailAdd{huhang3@mail2.sysu.edu.cn}
\emailAdd{chengjie@ncepu.edu.cn}
\emailAdd{guowl@ihep.ac.cn}
\emailAdd{wangw223@mail.sysu.edu.cn}

\abstract{
We explore the neutrino signals from proton decays catalyzed by GUT monopoles in the Sun. Three typical proton decay modes, $p \rightarrow e^+ + (\rho^0, \eta, \omega...) \rightarrow \pi^+$, $p \rightarrow \mu^+ K^0$ and $p \rightarrow \bar{\nu}_e \pi^+$, have been analyzed for the Super-Kamiokande experiment. The monopole-induced neutrinos arise from interactions and subsequent decays of the proton decay products. To obtain the neutrino energy spectra, we use the Geant4 software to simulate propagations of daughter particles in the highly-dense solar center. It is found that $K^0$ can produce a large amount of 236 MeV monoenergetic $\nu_\mu$ neutrinos through the charge exchange process $K^0 + p \rightarrow K^+ + n$ and the subsequent decay $K^+ \rightarrow \mu^+ \nu_\mu$ at rest. Based on this interesting feature, $p \rightarrow \mu^+ K^0$ can give the best discovery potential among three decay modes for most of the parameter space. In addition, we present the Super-Kamiokande sensitivities to the monopole flux for three proton decay modes.
}

\begin{document}
\maketitle
\flushbottom

%%%%%%%%%%%%%%%%%%%%%%%%%Section I %%%%%%%%%
\section{\label{sec:1} Introduction}

The existence of magnetic monopoles is an inherent prediction of all Grand Unified Theories (GUTs), since these theories will break spontaneously down to a subgroup containing the electromagnetic U(1) factor \cite{tHooft:1974kcl,Polyakov:1974ek}. The GUT monopole mass $M_M$ is related to the unification scale $M_{\rm GUT}$, and is the order of $10^{17} - 10^{18}$ GeV for $M_{\rm GUT} \sim 10^{16}$ GeV \cite{ParticleDataGroup:2020ssz}. However, no generally convincing experimental evidence of GUT monopoles has been found \cite{Burdin:2014xma,Patrizii:2015uea,Mavromatos:2020gwk}.  Rubakov \cite{Rubakov:1982fp} and Callan \cite{Callan:1982au} indicate that some certain GUT monopoles traversing the matter can catalyze nucleon decay reactions with a cross section of ordinary strong interactions, such as $p+ {\rm monopole} \rightarrow e^+ + \pi^0 + {\rm monopole}$. Significant efforts have been made to directly search for nucleon decays catalyzed by GUT monopoles, such as Kamiokande \cite{Kajita:1985aig}, Soudan \cite{Bartelt:1986cv}, IMB \cite{Becker-Szendy:1994kqw}, Baikal \cite{Baikal:1997kuo}, MACRO\cite{MACRO:2002iaq}, and IceCube \cite{IceCube:2014xnp}. In addition, GUT monopoles can be captured by celestial bodies, and the following catalyzed nucleon decays can generate heat and neutrinos. The indirect limits about the monopole abundance have been derived based on the heat observations from neutron stars \cite{Kolb:1982si} and white dwarfs \cite{Freese:1998es}, and the neutrino detections from the Sun \cite{Arafune:1983tr,Super-Kamiokande:2012tld}.

Monopole-catalyzed proton decays in the Sun can produce neutrinos through the following processes \cite{Arafune:1983tr}: $p \rightarrow e^+ \widetilde\pi$, $\widetilde\pi \rightarrow \pi^+$, $\pi^+ \rightarrow \mu^+ \nu_\mu$ and $\mu^+ \rightarrow e^+ \nu_e \,\bar\nu_\mu$, where the symbol $\widetilde\pi$ represents $\rho^0, \eta, \omega$ and so on. The Super-Kamiokande (Super-K) experiment has searched for these low energy neutrinos ($E_\nu \leq 53$ MeV) and given an upper limit on the monopole flux \cite{Super-Kamiokande:2012tld}. Based on the low energy SU(3)$\times$U(1) effective theory \cite{Bais:1982hm}, $p \rightarrow e^+ \widetilde\pi$ catalyzed by the GUT monopole is the dominant neutrino production mode, namely the branching ratio $B(p \rightarrow e^+ \widetilde\pi) \sim 1$. In Ref. \cite{Bais:1982hm}, the authors predict the hierarchy of proton decay modes: $B(p \rightarrow e^+ \widetilde\pi) : B(p \rightarrow \mu^+ K^0) \approx 1: (m_d/m_s)^2$ , where $(m_d/m_s)^2 \sim 1/400$ is from the short-distance current algebra masses or $(m_d/m_s)^2 \sim 1/2$ is from the long-distance constituent masses. Meanwhile, $p \rightarrow \bar{\nu}_e \pi^+$ and $p \rightarrow \mu^+ \pi$ are apparently forbidden  \cite{Bais:1982hm}. In fact, the SU(3)$\times$U(1) theory may allow $p \rightarrow \bar{\nu}_e \pi^+$ at a sufficiently short distance limit \cite{Houston:2018rvz}. In this case, $p \rightarrow \bar{\nu}_e \pi^+$ has a branching ratio of $B(p \rightarrow \bar{\nu}_e \pi^+) \sim 10^{-4}$ and directly produces a monoenergetic $\bar{\nu}_e$ of 459 MeV. It is found that $p \rightarrow \bar{\nu}_e \pi^+$ can give the greater discovery potential than $p \rightarrow \mu^+ K^0$ due to the smaller background and the larger neutrino cross section \cite{Houston:2018rvz}.

In this paper, we find that $K^0$ from $p \rightarrow \mu^+ K^0$ can also generate the high energy neutrino (236 MeV $\nu_\mu$) through the charge exchange process $K^0 + p \rightarrow K^+ + n$ and the subsequent decay $K^+ \rightarrow \mu^+ \nu_\mu$ at rest. Compared with the $K^0$ decay, $K^0 + p \rightarrow K^+ + n$ is non-negligible because of the considerable proton density at the center of the Sun. Here we shall explore the Super-K discovery potentials of $p \rightarrow e^+ \widetilde\pi$, $p \rightarrow \mu^+ K^0$ and $p \rightarrow \bar{\nu}_e \pi^+$ catalyzed by GUT monopoles in the Sun. This paper is organized as follows. Section \ref{sec:2} briefly introduce the monopole-catalyzed proton decays in the Sun. Then we discuss the neutrino production from three typical proton decay modes in section \ref{sec:3}, and give the corresponding neutrino fluxes at the Earth based on the Geant4 simulation. Section \ref{sec:4} shows the expected signal and background distributions in Super-K. In section \ref{sec:5}, we calculate the Super-K sensitivities to three typical proton decay modes and compare their discovery potentials. Finally, a conclusion will be given in section \ref{sec:6}.

%%%%%%%%%%%%%%%%%%%%%%%%%Section II %%%%%%%%%%%%%%%%%%
\section{\label{sec:2} Monopole-catalyzed proton decays in the Sun}

GUT monopoles can be produced in the very early Universe as stable topological defects during phase transitions via the Kibble mechanism \cite{Kibble:1976sj}. As the Universe expanded and cooled down, GUT monopoles could reach a speed of $\beta \sim 10^{-10}$. After the galaxy formation, they can be bound to our galaxy and be accelerated by the galactic magnetic field to $\beta \sim 10^{-3} \sqrt{10^{17} \,{\rm GeV}/M_M}$ for the monopole mass $M_M \gtrsim 10^{11}$ GeV \cite{ParticleDataGroup:2020ssz}. Comparing the energy loss rate with the regeneration rate of the galactic magnetic field, Parker obtained a bound on the monopole flux  \cite{Parker:1970xv, Turner:1982ag}:
\begin{equation}
   F_M <
   \left\{
             \begin{array}{lr}
             10^{-15} \, {\rm cm^{-2} \, sr^{-1} \, s^{-1}},   &  M_M \lesssim 10^{17} \, {\rm GeV},  \\
             10^{-15} \left( \frac{M_M}{10^{17} {\rm GeV}} \right) {\rm cm^{-2} \, sr^{-1} \, s^{-1}},   &  M_M \gtrsim 10^{17} \, {\rm GeV} .
             \end{array}
    \right.
\end{equation}
The intergalactic GUT monopoles are isotropic due to the acceleration process.

Some GUT monopoles passing through the Sun can lose enough energy and are captured by the Sun. The dominant energy loss mechanism comes from the electronic interactions between GUT monopoles and electrons in the Sun \cite{Frieman:1985dv}. The total number of monopoles trapped by the Sun is given by \cite{Arafune:1983tr,Super-Kamiokande:2012tld}
\begin{eqnarray}
    N_M & = & 4 \pi F_M \pi R_\odot^2 \left [ 1 + \frac{\beta^2_{\rm esc}}{\beta^2} \right] \epsilon (M_M,\beta, g) t_{\odot}  \nonumber \\ 
    & = & 2.8 \times 10^{25}   \frac{F_M}{10^{-15} \, {\rm cm^{-2} \, sr^{-1} \, s^{-1}}}  \left [ 1 + \frac{\beta^2_{\rm esc}}{\beta^2} \right] \epsilon (M_M,\beta, g)  , 
    \label{eq2}
\end{eqnarray}
where the solar radius $R_{\odot}=7.0 \times 10^{10}$ cm and the solar age $t_{\odot} = 4.6 \times 10^{9}$ yr. The term in the bracket accounts for the focusing effect of the solar gravitational field, and $\beta_{\rm esc} = 2 \times 10^{-3}$ is the escape velocity at the Sun surface. $\epsilon (M_M,\beta,g)$ describes the capture fraction of all monopoles that enter the Sun can be captured. It can be derived from the monopole stopping power \cite{Frieman:1985dv,Ahlen:1996ax}, which depends on the monopole mass $M_M$, velocity $\beta$ and magnetic charge $g$. The capture fraction $\epsilon (M_M,\beta,g)$ will increase and approach 1 as the monopole mass and velocity decrease. In Refs. \cite{Frieman:1985dv,Ahlen:1996ay},  $\epsilon (M_M,\beta,g)$ has been numerically calculated by solving the monopole motion equation. For a typical GUT monopole with $M_M = 10^{16}$ GeV, $\beta = 10^{-3}$ and the Dirac magnetic charge, we can obtain $\epsilon = 0.48$ from Ref. \cite{Ahlen:1996ay} and give $N_M = 6.7 \times 10^{25}$ when $F_M$ takes the Parker bound $1 \times 10^{-15} \, {\rm cm^{-2} \, sr^{-1} \, s^{-1}}$.

Once these GUT monopoles stop in the Sun, they will quickly fall to the solar center, and their distribution depends on the support mechanism against gravity \cite{Frieman:1985dv}. When monopoles are supported by their own thermal pressure, the distribution radius is order of $10^2$ cm. In this case, the monopole-antimonopole annihilation can drastically reduce the number of captured monopoles. If the solar center has a magnetic field of several hundred Gauss, it can support monopoles to a distance of $\sim 10^7$ cm, and prevent the annihilation \cite{Frieman:1985dv}. In the following analysis, we assume that the trapped monopoles are uniformly distributed in the solar core of radius $r_M \sim 10^7$ cm and the monopole-antimonopole annihilation is negligible.

According to the Rubakov-Callan effect \cite{Rubakov:1982fp,Callan:1982au}, the captured monopoles can catalyze proton decays in the Sun. The cross section $\sigma_{R}$ of the catalysis process behaves as \cite{Rubakov:1982fp,Callan:1982au,Arafune:1983uz}
\begin{equation}
   \sigma_{R} =  \frac{\sigma_0}{\beta_{\rm rel}} \cdot F(\beta_{\rm rel}), 
\end{equation}
where $\sigma_0$ is estimated to be the order of the hadronic cross sections. The relative velocity between monopoles and protons in the Sun may be taken from the Hydrogen thermal velocity, $\beta_{\rm rel} = \sqrt{2 \, T_\odot / m_p} = 1.7 \times 10^{-3}$ with the solar central temperature $T_\odot = 1.544 \times 10^7$ K \cite{Vinyoles:2016djt}. $F(\beta_{\rm rel})$ is a correction factor of the catalysis process for slowly moving monopoles in the matter. Based on $F(\beta_{\rm rel})$ listed in Table I of Ref. \cite{Arafune:1983uz}, the Hydrogen element with $F(\beta_{\rm rel}) = 0.17/\beta_{\rm rel}$ gives a dominant contribution to monopole-catalyzed proton decays in the Sun, and contributions from other elements are negligible. The monopole-catalyzed proton decay rate in the Sun is given by \cite{Arafune:1983sk}
\begin{equation}
    f_p = \frac{\rho_H}{m_p} \, \beta_{\rm rel} \, \sigma_R \, N_M = 9.8 \times 10^{10} \, \frac{\sigma_0}{1 \, {\rm mb}} N_M \, {\rm s^{-1}}, 
    \label{eq4}
\end{equation}
where a fixed Hydrogen mass density $\rho_H =53.9 \, {\rm g \, cm^{-3}}$ \cite{Vinyoles:2016djt} has been used since these trapped monopoles are confined to a very small region $r_M/R_\odot < 0.001$.

%%%%%%%%%%%%%%%%%%%%%%%%%Section III  %%%%%%%%%%%%%%%%%

\section{\label{sec:3} Neutrino fluxes at the Earth}

Monopole-catalyzed proton decays in the Sun can produce neutrinos through subsequent decays of daughter particles. On the other hand, we should consider interactions of final state mesons in the Sun, such as the $\pi^+$ absorption and the $K^0$ charge exchange process, which can significantly change the produced neutrino fluxes and energy spectra. In addition, these monopole-induced neutrinos will undergo the neutrino oscillation from the solar center to neutrino detectors. The neutrino flux at the surface of the Earth can be written as 
\begin{equation}
    \frac{d\Phi_{\nu_\alpha}}{dE_{\nu}} =  \frac{f_p}{4\pi R^2}  B \, \sum_{l=(e, \mu)} Y_{\nu_l} \, f_{\nu_l}(E_\nu) \, P_{\nu_l \rightarrow \nu_\alpha} \,,  
\label{flux}
\end{equation}
where $\nu_\alpha = (\nu_e,\nu_\mu,\nu_\tau)$ and $R = 1.5 \times 10^{13}$ cm is the Earth-Sun distance. For the branching ratios $B$ of three typical proton decay modes, we take 
\begin{equation}
\begin{aligned}
 B(p \rightarrow e^+ \widetilde\pi \rightarrow \pi^+) & \equiv B(p \rightarrow e^+ \widetilde\pi)  f_{\pi^+} \approx   0.5, \\ 
 B(p \rightarrow \mu^+ K^0) & \approx {m_d^2}/{m_s^2} \sim 1/400-1/2, \\
 B(p \rightarrow \bar{\nu}_e \, \pi^+) & \approx {m_p^2}/{m_W^2} \sim 10^{-4}, 
    \label{B}
\end{aligned}
\end{equation}
where $B(p \rightarrow \mu^+ K^0)$ and $B(p \rightarrow \bar{\nu}_e \pi^+)$ come from the theoretical predictions \cite{Bais:1982hm,Houston:2018rvz}, and $B(p \rightarrow e^+ \widetilde\pi \rightarrow \pi^+)$ is the same with the Super-K assumption  \cite{Super-Kamiokande:2012tld}. Note that the effective branching ratio $B(p \rightarrow e^+ \widetilde\pi \rightarrow \pi^+)$ is defined as $B(p \rightarrow e^+ \widetilde\pi)$ multiplied by the average $\pi^+$ production rate $f_{\pi^+}$ from a proton decay of $p \rightarrow e^+ \widetilde\pi$. We have performed the simulations of $p \rightarrow e^+ \widetilde\pi$, such as $p \rightarrow e^+ \eta$, $p \rightarrow e^+ \rho^0$ and so on. These channels only produce low energy neutrinos from the secondary $\pi^+$ decay other than high energy (monoenergetic) neutrinos. Since analyses of these channels are same with the $\pi^+$ case, we directly simulate the $\pi^+$ propagation in the Sun. The only difference between these channels and the $\pi^+$ channel is the neutrino yield. So their results can be obtained by a scaled factor from the $\pi^+$ case. Here we have conservatively assumed $B(p \rightarrow e^+ \widetilde\pi \rightarrow \pi^+) \approx 0.5$. The $p \rightarrow \bar{\nu}_\mu K^+$ can produce high energy neutrinos. However, the branching ratio of this channel is about $10^{-4}$ of $p \rightarrow \mu^+ K^0$ based on the discussions in Refs. \cite{Bais:1982hm} and \cite{Houston:2018rvz}. Therefore, we only analyze the three typical proton decay modes in Table \ref{neutrino:product-rate}. In Eq. (\ref{flux}), $P_{\nu_l \rightarrow \nu_\alpha}$ is the oscillation probability of $\nu_l$ from the Sun center to $\nu_\alpha$ at the Earth surface. In the following analysis, $P_{\nu_l \rightarrow \nu_\alpha}$ will be taken from Fig. 6 of Ref. \cite{Baratella:2013fya}, where $\theta_{13} = 8.8^\circ$ differs slightly from the current best-fit value \cite{ParticleDataGroup:2020ssz}. Assuming the normal hierarchy, we get $P_{\nu_\mu-\nu_e}(236 \, \rm{MeV})={0.48}$, $P_{\bar\nu_e-\bar\nu_e(\bar\nu_\mu)}(459\, \rm{MeV})={0.67 \, (0.23)}$. In the low energy range, the oscillation probability of $P_{\bar\nu_\mu-\bar\nu_e}$ is insensitive to the neutrino energy. It is convenient for us to take a fixed $P_{\bar\nu_\mu-\bar\nu_e} = 0.17$ for 20 MeV $\leq E_\nu \leq 53$ MeV.

\begin{table}
\setlength{\belowcaptionskip}{0.2cm}
\centering
\caption{Neutrino yields $Y_{\nu_l}$ of three proton decay modes in the solar center from the Geant4 simulation. For the monoenergetic spectrum, the value in the parenthesis means the neutrino energy.}
\begin{tabular}{l|c|c|c|c|c|c|c} \hline  \hline
&\multicolumn{4}{c|}{Continuous Spectrum} &\multicolumn{3}{c}{Monoenergetic Spectrum} \\ \hline
Decay mode    & $Y_{\nu_{e}}$   & $Y_{\bar{\nu}_e}$  & $Y_{\nu_\mu}$   & $Y_{\bar{\nu}_\mu}$  & $Y_{\nu_\mu}(30)$ & $Y_{\nu_\mu}(236)$ &$Y_{\bar{\nu}_e}(459)$ \\    \hline
$p \rightarrow e^+ \widetilde\pi \rightarrow \pi^+ $    & 0.86   & -      & - & 0.86   &0.86  & - & - \\ \hline
$p \rightarrow \mu^+  K^0 $   &1.79    & $<$0.01      &0.01   & 1.77   & 0.55  & 0.21 & - \\ \hline
$p \rightarrow \bar{\nu}_e \, \pi^{+}$         &0.72    &-        &-   & 0.72  & 0.72 & - & 1.0 \\
  \hline \hline
 \end{tabular}
\label{neutrino:product-rate}
\end{table}

In Eq. (\ref{flux}), $Y_{\nu_l}$ and $f_{\nu_l}(E_\nu)$ describe the $\nu_l$ yield and the normalized energy spectrum from proton decays in the Sun, respectively. In order to determine $Y_{\nu_l}$ and $f_{\nu_l}(E_\nu)$ for three typical proton decay modes, we use the Geant4 version 10.7.patch-03 \cite{GEANT4:2002zbu} with the FTFP$\_$BERT physics list to simulate propagations of daughter particles $\pi^+, \mu^+$ and $K^0$ with corresponding momenta in the core of the Sun. In this simulation, the solar central density $\rho = 148.9 \,{\rm g/cm^3}$ and the composition of dominant elements from the AGSS09 solar model have been used \cite{Vinyoles:2016djt}. The neutrino production yields are listed in Table \ref{neutrino:product-rate}. Due to inelastic reactions with nucleons, about $28\%$ $\pi^+$ with a momentum of 459 MeV can not produce low energy neutrinos ($E_\nu \leq 53$ MeV). As the $\pi^+$ momentum goes down to 100 MeV, $Y_{\nu_l}$ will increase from 0.72 to 0.99. Since the $\pi^+$ momentum distribution is model-dependent, we here take the average value 0.86 of 0.72 and 0.99 for $p \rightarrow e^+ \widetilde\pi \rightarrow \pi^+$. $K^0$ can also produce low energy neutrinos through $K^0_S \rightarrow \pi^+ \pi^-$ and the $K^0_S$ regeneration from $K^0_L$. In addition, high energy neutrinos ($E_\nu \geq 200$ MeV) can be expected from the direct decay of $K^0_L \rightarrow \pi^\pm e^\mp \bar{\nu}_e/\nu_{e}$. The corresponding probability is estimated to be about $2.0 \times 10^{-6}$ \cite{Arafune:1983sk}. Therefore, one usually believes that high energy neutrinos are negligible for the decay mode $p \rightarrow \mu^+ K^0$. However, we find that a $K^0$ with a momentum of 326 MeV can averagely produce 0.21 monoenergetic 236 MeV $\nu_\mu$, $Y_{\nu_\mu}(236) = 0.21$. This is because that the charge exchange process $K^0 + p \rightarrow K^+ + n$  is no longer negligible compared with the $K^0_S$ decay in the highly-dense center of the Sun. Note that different solar models only slightly change the results in Table \ref{neutrino:product-rate}. Keeping the mass fractions of elements unchanged, we use $\rho = 100.0 \,{\rm g/cm^3}$ to calculate the neutrino yields for 459 MeV $\pi^+$ and 326 MeV $K^0$. It is found that $Y_{\nu_e} = 0.72$ from 459 MeV $\pi^+$ is same with that in the standard $148.9 \, {\rm g/cm^3}$ case. For the 326 MeV $K^0$, the $Y_{\nu_\mu} (236)$ in the 148.9 ${\rm g/cm^3}$ and 100.0 ${\rm g/cm^3}$ cases are 0.21 and 0.19, respectively. In Table \ref{reaction}, we list the percentages of the $K^0$ last reactions in the solar center from the Geant4 simulation. It is worthwhile to stress that the Geant4 considers the Hypernucleus production process \cite{Wright:2015xia}, such as $K^0 + ^{4}{\rm He} \rightarrow ^{4}_{\Lambda}{\rm H} + \pi^+$. Based on the $K^+$ production percentage of $32.3\%$ and the $K^+ \rightarrow \mu^+ \nu_\mu$ branching ratio of $63.56\%$, one can easily estimate $Y_{\nu_\mu}(236) = 0.21$. To test this result, we have also used the FLUKA (version 2021) \cite{Bohlen:2014buj} to simulate the $K^0$ propagation in the Sun and obtained $Y_{\nu_\mu}(236) = 0.22$. The corresponding percentages of the $K^0$ last reactions in the solar center have been listed in Table \ref{reaction}. Except for the Hypernucleus production process, there are no significant differences between the Geant4 and FLUKA simulations. The conclusion that $K^0$ in the solar center can produce a large amount of 236 MeV $\nu_\mu$ neutrinos is reliable. Here we use the Geant4 results for the following analyses.

\begin{table}[]
\setlength{\belowcaptionskip}{0.2cm}
\centering
\caption{Percentages of the $K^0$ last reactions in the solar center from the Geant4 and FLUKA simulations. For the $\Lambda$ Hyperon, $\Sigma$ Hyperon and Hypernucleus production processes, the accompanying $\pi^+$ yields multiplied by a factor of 100 are given in the parenthesis.}
\begin{tabular}{c|c|c|c|c|c|c} 
\hline \hline
$K^0$ reactions  &  $K^0_S$ decay    & $K^0_L$ decay  & $K^+/K^-$    & $\Lambda$ Hyperon& $\Sigma$ Hyperon & Hypernucleus   \\ \hline
 Geant4  & 18.6 & < 0.1   & 32.3/2.9     & 24.7 (22.6)    & 13.5 (6.4)    & 8.0 (4.0)    \\ \hline
 FLUKA  &  40.3  & 0.2   &  34.7/0.3     & 10.4 (8.7) &   14.1 (6.8)  &  0      \\ 
  \hline \hline
      \end{tabular}
    \label{reaction}
\end{table}

\begin{figure}
    \centering
    \includegraphics[width=0.48\linewidth]{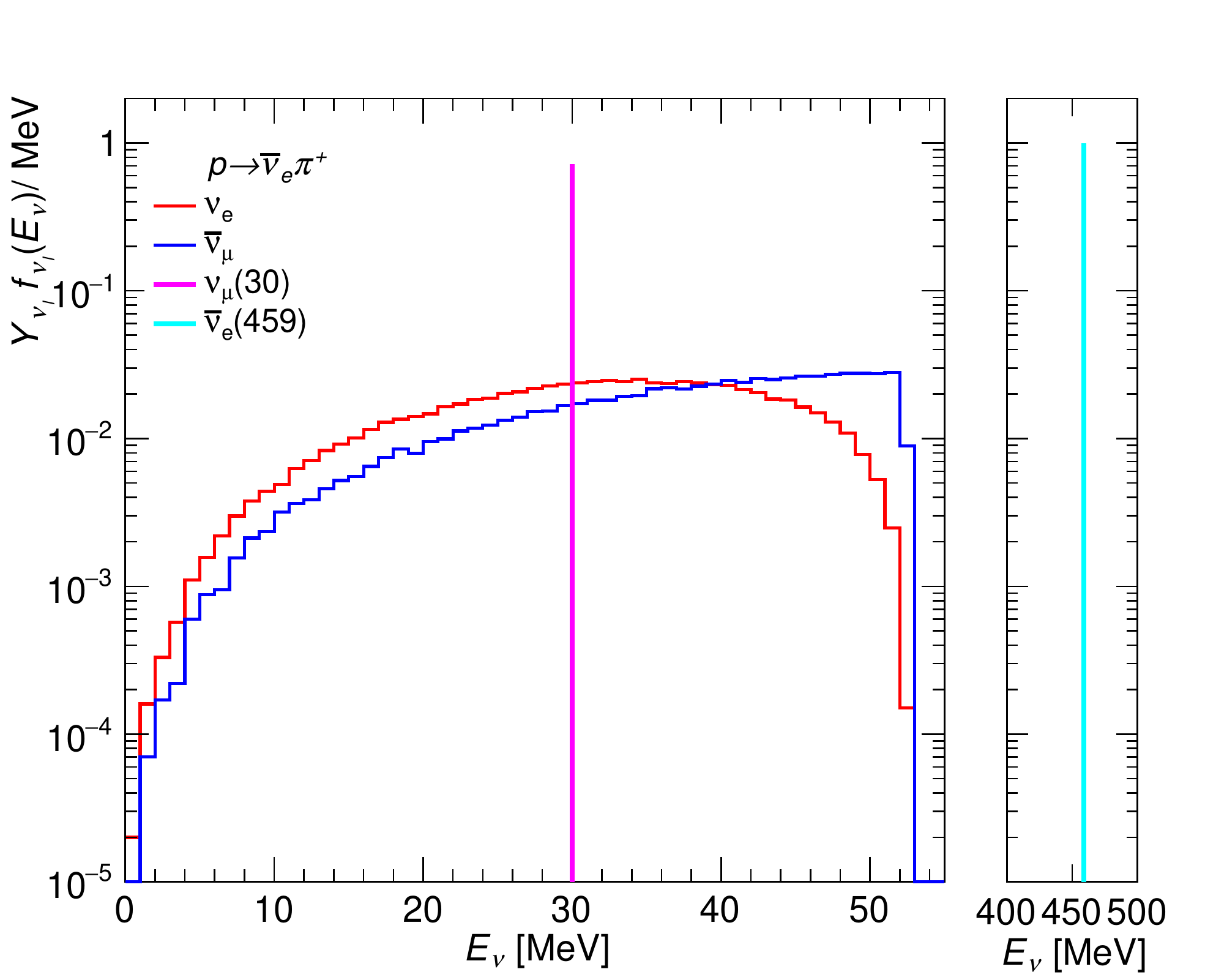}
    \includegraphics[width=0.48\linewidth]{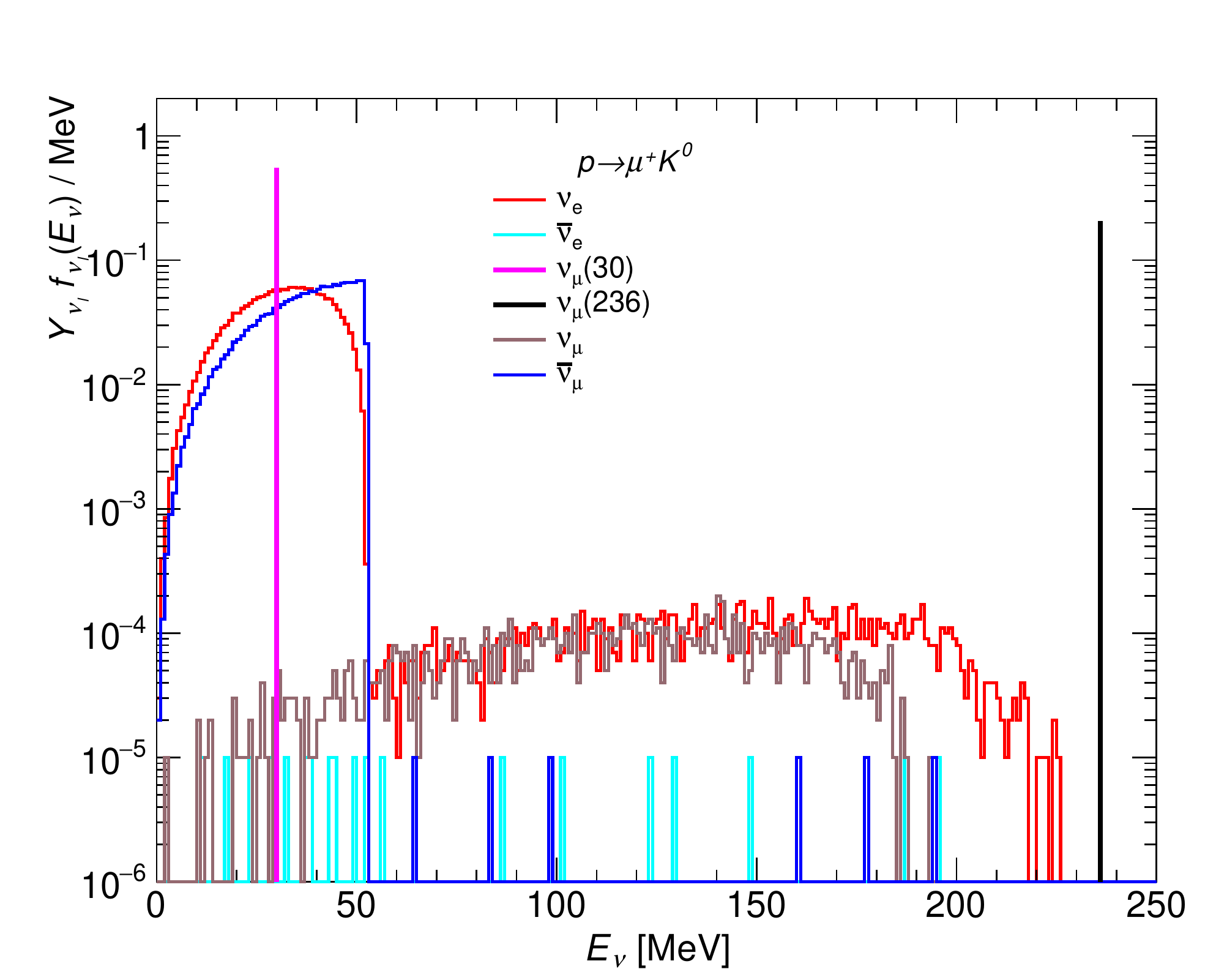}
    \caption{ The neutrino energy spectra $Y_{\nu_l} f_{\nu_l}(E_\nu)$ evaluated by the Geant4 simulation for monopole-catalyzed proton decays $p \rightarrow \bar{\nu}_e \, \pi^{+}$ (left) and $p \rightarrow \mu^+  K^0 $ (right) in the Sun.} 
    \label{neutrios:spectra}
\end{figure}

In Fig. \ref{neutrios:spectra}, we plot the $\nu_l$ energy spectra $Y_{\nu_l} f_{\nu_l}(E_\nu)$ from the Geant4 simulations of $p \rightarrow \bar{\nu}_e \pi^+$ and $p \rightarrow \mu^+ K^0$ in the Sun. For the decay mode $p \rightarrow \bar{\nu}_e \pi^+$, low energy $\nu_\mu$ (30 MeV), $\nu_e$ and $\bar\nu_\mu$ neutrinos are produced through the two-step decay process: $\pi^+ \rightarrow \mu^+\nu_\mu$ and $\mu^+ \rightarrow e^+ \nu_e \, \bar\nu_\mu$. The obtained $\nu_e$ and $\bar\nu_\mu$ energy spectra are well consistent with the theoretical calculation of the $\mu^+$ decay at rest \cite{Guo:2018sno}. $f_{\nu_\mu}(E_\nu) \simeq \delta(E_\nu - 30\, {\rm MeV})$ implies that the $\pi^+$ decaying in the flight is negligible due to the large density in the solar center. For $p \rightarrow e^+ \widetilde\pi$, $f_{\nu_l}(E_\nu \leq 53 {\rm MeV} )$ from $p \rightarrow \bar{\nu}_e \pi^+$ will be used. For the decay mode $p \rightarrow \mu^+ K^0$, low energy neutrinos $\nu_e$ and $\bar{\nu}_\mu$ have four dominant sources in terms of Table \ref{reaction}: the final state $\mu^+$, $\pi^+$ from $K^0_S \rightarrow \pi^+ \pi^-$, $\mu^+$ and $\pi^+$ from $K^+$ decays, and the accompanying $\pi^+$ in the $\Lambda$ Hyperon, $\Sigma$ Hyperon and Hypernucleus production processes. In addition, $\Sigma^+$ with a $6.1 \%$ production possibility can also decay a $\pi^+$ through $\Sigma^+ \rightarrow n \, \pi^+$. Based on the percentages and $\pi^+$ yields in Table \ref{reaction}, and the decay branching ratios of $K^0_S$, $K^+$ and $\Sigma^+$, one can simply estimate the low energy neutrino yields $Y_{\nu_l}$ for the $p \rightarrow \mu^+  K^0 $ mode, which are basically consistent with the full simulation results in Table \ref{neutrino:product-rate}. As shown in the right panel of Fig. \ref{neutrios:spectra}, $K^0$ can produce a small amount of $\nu_e$, $\bar{\nu}_e$, $\nu_\mu$ and $\bar{\nu}_\mu$ in the energy range of 53 MeV $\leq E_\nu \leq$ 230 MeV. We find that the antineutrino spectra are far smaller than the neutrino spectra in this region. It means that these neutrinos basically arise from the $K^+$ decay processes $K^+ \rightarrow \pi^0  e^+ \nu_e$ and $K^+ \rightarrow \pi^0  \mu^+  \nu_\mu$, rather than from $K^0_L \rightarrow \pi^\pm e^\mp \bar{\nu}_e/\nu_{e}$ and $K^0_L \rightarrow \pi^\pm \mu^\mp \bar{\nu}_\mu/\nu_{\mu}$. This result agrees with the percentage $< 0.1\%$ of $K^0_L$ decays in Table \ref{reaction}. Because of $Y_{\nu_l} \sim 1\%$ in this range, we do not consider their contributions in this paper.

\section{\label{sec:4} Neutrino detections in Super-K }

Here we shall discuss the neutrino signals in the Super-K detector from monopole-catalyzed proton decays. The expected $\nu_\alpha$ event number from the charged current (CC) interaction can be calculated by
\begin{equation}
    N_{\nu_\alpha} = N_{\rm target} \, T \, \int \frac{d\Phi_{\nu_\alpha}}{dE_{\nu}} \, \sigma_{\nu_\alpha}(E_\nu) \,  dE_{\nu},
\label{Nv}
\end{equation}
where $N_{\rm target}$ is the target number and the exposure time $T = 7.8$ years will be taken for the comparison with the previous result in Ref. \cite{Super-Kamiokande:2012tld}. One can quickly obtain $1.5 \times 10^{33}$ free protons and $7.5 \times 10^{32}$ water molecules from the Super-K fiducial mass of 22.5 ktons. Here we only consider the inverse beta decay (IBD) reaction for the decay mode $p \rightarrow e^+ \widetilde\pi$ since its event rate is far larger than the electron elastic scattering \cite{Super-Kamiokande:2012tld}. The IBD cross section $\sigma_{\bar\nu_e}(E_\nu)$ can be found in Ref. \cite{Strumia:2003zx}, which is the order of $10^{-41}-10^{-40}$ cm$^2$ in the energy range of $20 \, {\rm MeV} \leq E_\nu \leq 53$ MeV. For high energy neutrinos, we evaluate the CC cross sections per water molecule from the GENIE database \cite{Andreopoulos:2009rq}, and obtain $\sigma_{\nu_e}(\rm{236 \, MeV}) = 2.51\times 10^{-38} \,  \rm{cm^2}$, $\sigma_{\bar{\nu}_e}(\rm{459 \,  MeV}) = 2.16 \times 10^{-38} \,  \rm{cm^2}$ and $\sigma_{\bar{\nu}_\mu}(\rm{459 \,  MeV}) = 2.00 \times 10^{-38}  \, \rm{cm^2}$. It is clear that they are much larger than the low energy IBD cross section. So only the high energy neutrino events will be analyzed for the proton decay modes $p \rightarrow \bar{\nu}_e \pi^+$ and $p \rightarrow \mu^+ K^0$ in the following parts. With the help of Eqs. (\ref{flux}) and (\ref{Nv}), we can express the expected $\nu_\alpha$ event number as
\begin{equation}
\begin{aligned}
    N_{\bar{\nu}_e}(\leq 53 \, \rm{MeV}) & = 5.29 \times  \frac{f_p/(4 \pi R^2) }{\rm 1 \, cm^{-2} \, s^{-1}} \times B(p \rightarrow e^+ \widetilde\pi \rightarrow \pi^+), \\ 
    N_{\nu_e}(236 \, \rm{MeV}) & = 4.60 \times 10^2 \times  \frac{f_p/(4 \pi R^2) }{\rm 1 \, cm^{-2} \, s^{-1}} \times B(p \rightarrow \mu^+ K^0), \\ 
    N_{\bar{\nu}_e}(459 \, \rm{MeV}) & =  2.68 \times 10^3 \times  \frac{f_p/(4 \pi R^2)}{\rm 1 \, cm^{-2} \, s^{-1}}  \times B(p \rightarrow \bar{\nu}_e \, \pi^+) , \\  
    N_{\bar{\nu}_\mu}(459 \, \rm{MeV}) & = 8.52 \times 10^2  \times  \frac{f_p/(4 \pi R^2)}{\rm 1 \, cm^{-2} \, s^{-1}}  \times B(p \rightarrow \bar{\nu}_e \, \pi^+),
    \label{Nsignal}
\end{aligned}
\end{equation}
for the 176 kton$\cdot$year exposure of the Super-K detector.

\subsection{\label{sec:pi} Analysis of $p \rightarrow e^+ \widetilde\pi$}

\begin{figure}
    \centering
    \includegraphics[width=0.47\linewidth]{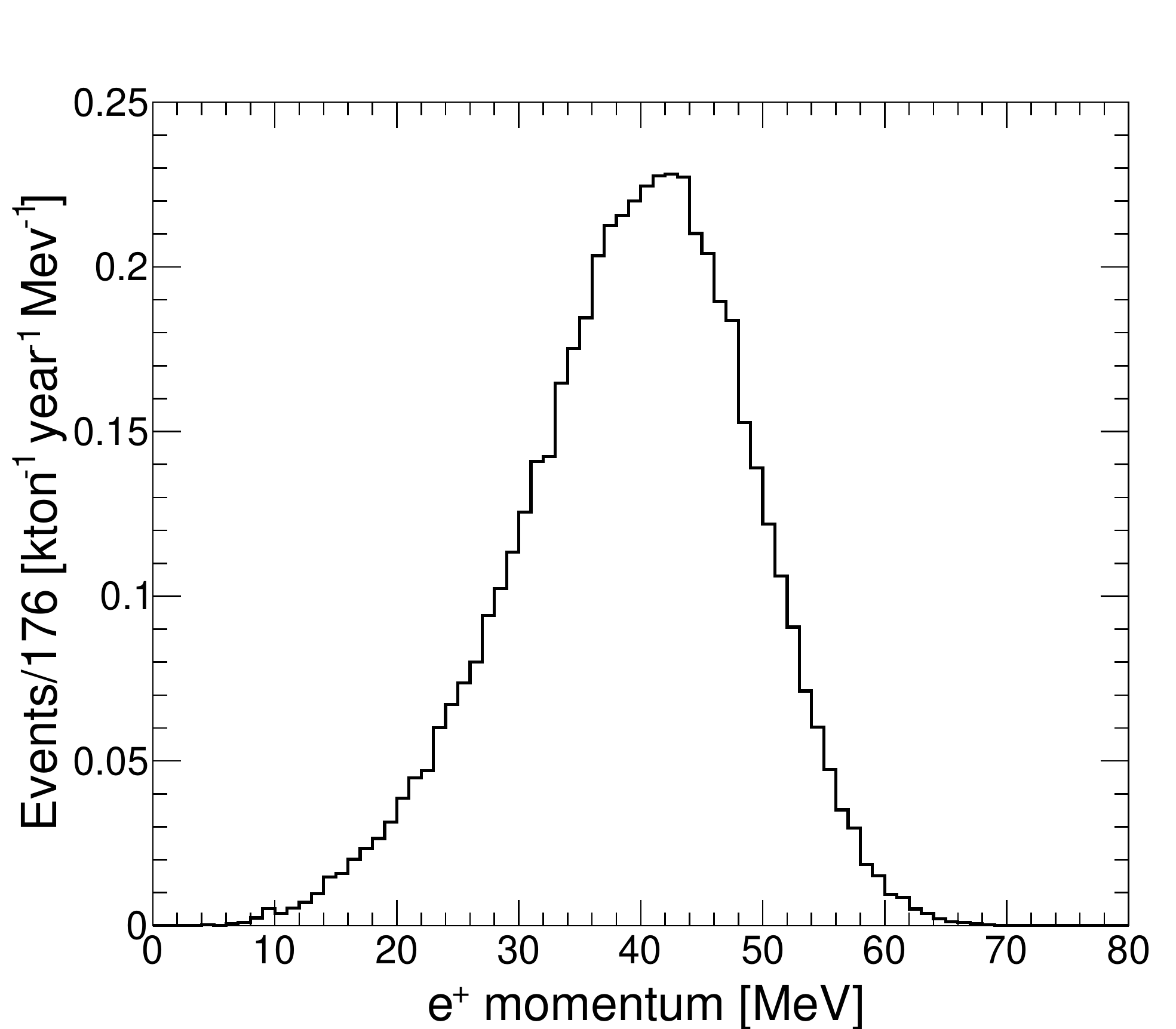}
    \caption{The expected distribution of the IBD events as a function of the $e^+$ momentum with $f_p/(4\pi R^2) B(p \rightarrow e^+ \widetilde\pi \rightarrow \pi^+) = 1 \, \rm{cm^{-2} s^{-1}}$ and the Super-K 176 kton$\cdot$year exposure.} 
    \label{ibd}
\end{figure}

The Super-K experiment has performed the search of monopole-induced neutrinos from the decay mode $p \rightarrow e^+ \widetilde\pi$ \cite{Super-Kamiokande:2012tld}. In order to compare the significances of three proton decay modes, we here do not cite the Super-K result, and shall adopt an uniform method to estimate their sensitivities. Here we use GENIE version 3.0.2 with the G18$\_$02a$\_$02$\_$11a model set \cite{Andreopoulos:2009rq} to generate the predicted momentum spectrum of positrons from the IBD reaction $\bar{\nu}_e + p \rightarrow e^+ + n$. The inputting $\bar{\nu}_e$ spectrum originates from the theoretical $\bar{\nu}_\mu$ spectrum of the $\mu^+$ decay at rest. The expected distribution of IBD events as a function of the $e^+$ momentum $p_e$ is shown in Fig. \ref{ibd}. It has been normalized to the number of 5.29 in Eq. (\ref{Nsignal}). Note that the momentum resolution of $0.6+2.6/\sqrt{p_e ({\rm GeV})}\%$ \cite{Super-Kamiokande:2005mbp} has been included in this figure. To reduce backgrounds from the spallation products and solar neutrinos, we take the momentum cut 20 MeV $\leq p_e \leq$ 55 MeV, which is basically consistent with the selection condition [19-55] MeV on the reconstructed event energy \cite{Super-Kamiokande:2012tld}. Considering the above momentum range and other cuts listed in Table 1 of Ref. \cite{Super-Kamiokande:2012tld}, we may derive the signal efficiency $\varepsilon = 0.82$. In this case, the expected background number $N_{\rm bkg} \simeq 300$ and the observed event number $N_{\rm obs} = 317$ can be found in Ref. \cite{Super-Kamiokande:2012tld}. The background events are dominantly caused by atmospheric neutrino interactions, such as the decay electrons from invisible muons, the final state electrons and the multiple de-excitation $\gamma$-rays from residual nuclei.

\subsection{\label{sec:K0} Analysis of $p \rightarrow \mu^+ K^0$}

\begin{figure}
    \centering
    \includegraphics[width=0.47\linewidth]{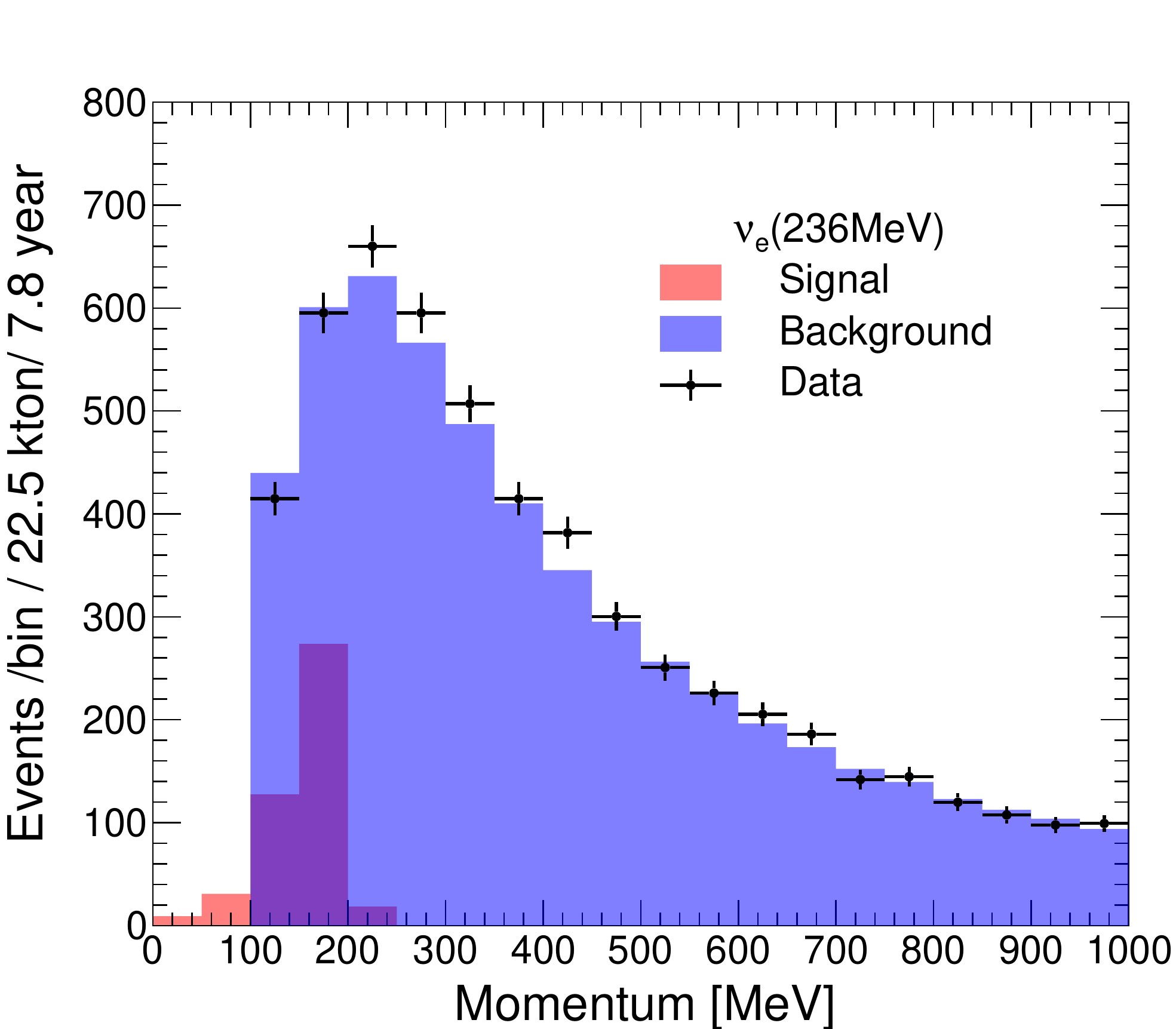}
    \caption{ The expected momentum distribution of electrons from the 236 MeV $\nu_e$ CC interaction assuming $f_p/(4\pi R^2) B(p \rightarrow \mu^+ K^0) = 1 \, \rm{cm^{-2} s^{-1}}$. The background and observed momentum distributions come from Ref. \cite{Super-Kamiokande:2014pqx}, and are scaled to the level of 176 kton$\cdot$year exposure. } 
    \label{nue-236}
\end{figure}

For the 236 MeV $\nu_\mu$ produced by $p \rightarrow \mu^+ K^0$ in the Sun, the Super-K experiment can observe $\nu_\mu$ and $\nu_e$ CC events due to the neutrino oscillation possibility $P_{\nu_\mu-\nu_e}(236\rm{MeV})={0.48}$.  We use the GENIE generator to simulate the $\nu_\mu$ and $\nu_e$ CC interactions in the water. It is found that the average momentum of the final state $\mu^-$ is about 129 MeV. Since the momentum threshold of the muon Cherenkov radiation is 120 MeV, the Super-K can not effectively use the 236 MeV $\nu_\mu$ CC events. In addition, the Super-K does not give the experimental data with the momentum smaller than 200 MeV for the $\mu$-like events in Ref. \cite{Super-Kamiokande:2014pqx}. So we only calculate the $\nu_e$ contribution for the $p \rightarrow \mu^+ K^0$ analysis. For the $\nu_e$ CC interaction, we use the GENIE to generate the expected momentum distribution of electrons. Then it is normalized to the number of 460 in Eq. (\ref{Nsignal}), which can be numerically calculated from Eqs. (\ref{flux}) and (\ref{Nv}). The momentum distribution of electrons from the 236 MeV $\nu_e$ CC interaction is shown in Fig. \ref{nue-236}, where a momentum resolution of $0.6+2.6/\sqrt{p_e ({\rm GeV})}\%$ has been used.  It is clear that most events give $p_e \geq$ 100 MeV. The momentum distributions of background and observed events in the Super-K detector have also been plotted in Fig. \ref{nue-236}. These data come from Ref. \cite{Super-Kamiokande:2014pqx}, and are scaled to the level of 176 kton$\cdot$year exposure. Comparing the signal and background distributions, we set the selection condition 100 MeV $\leq p_{e} \leq$ 200 MeV to reduce atmospheric neutrino backgrounds and increase the discovery potential of $p \rightarrow \mu^+ K^0$. The corresponding signal efficiency $\varepsilon$, background number $N_{\rm bkg}$ and observed number $N_{\rm obs}$ have been summarized in Table \ref{summary}. Here we do not consider the direction cut about the final state electrons since their directions from the GENIE simulation are nearly isotropic at this energy.

\subsection{\label{sec:veb} Analysis of $p \rightarrow \bar{\nu}_e \, \pi^+$}

\begin{figure}
    \centering
    \includegraphics[width=0.47\linewidth]{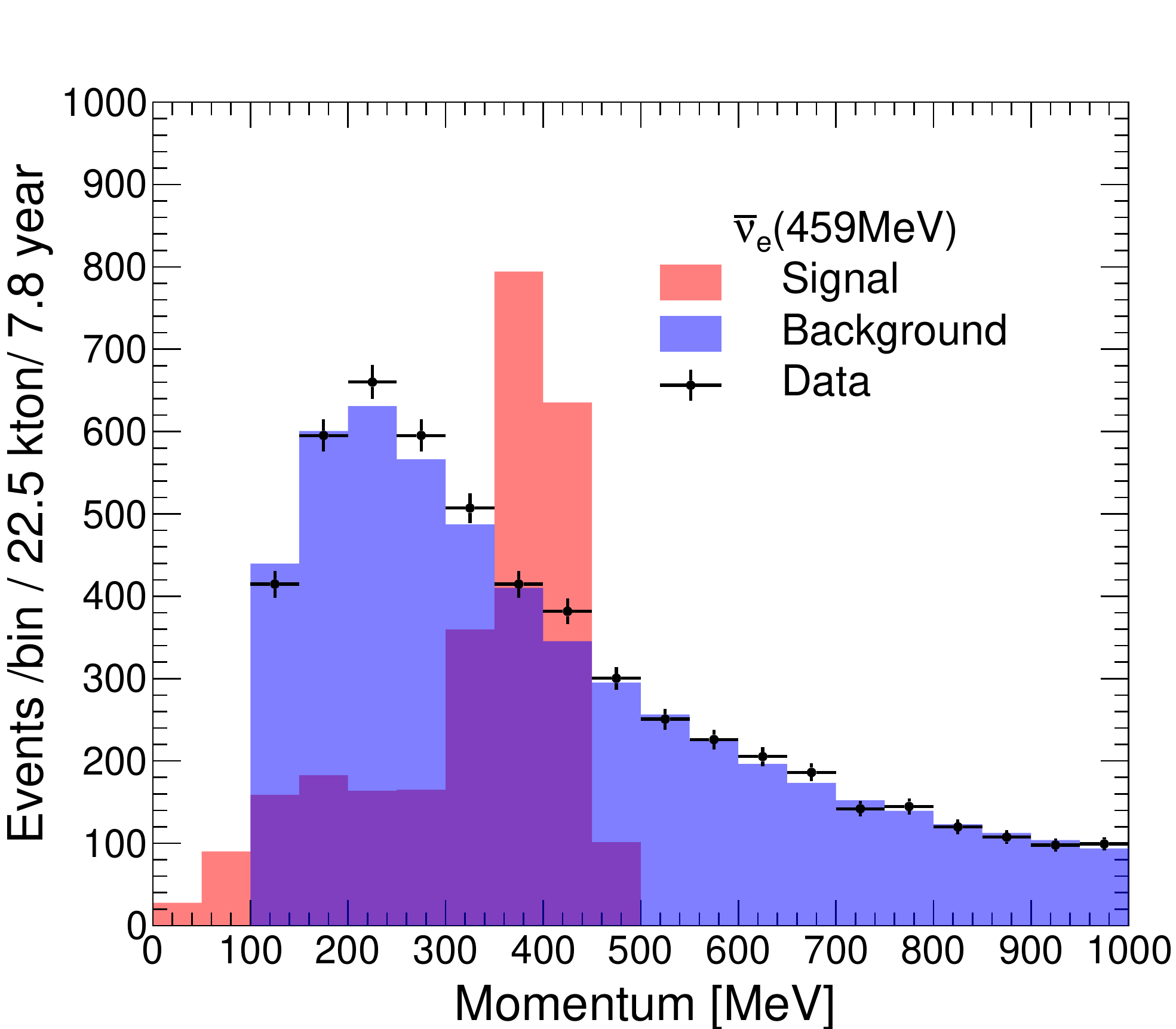}
    \includegraphics[width=0.47\linewidth]{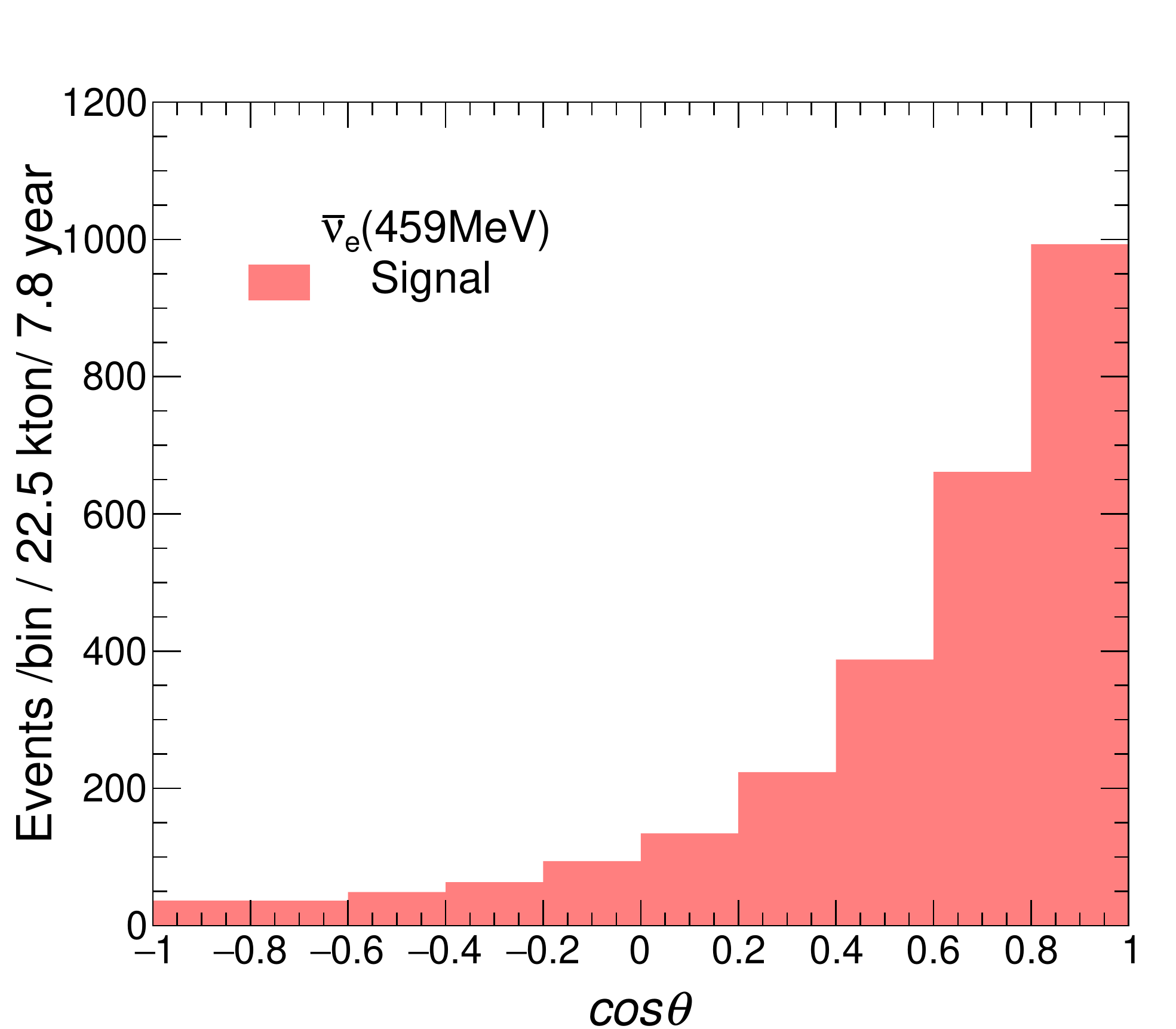}
    \includegraphics[width=0.47\linewidth]{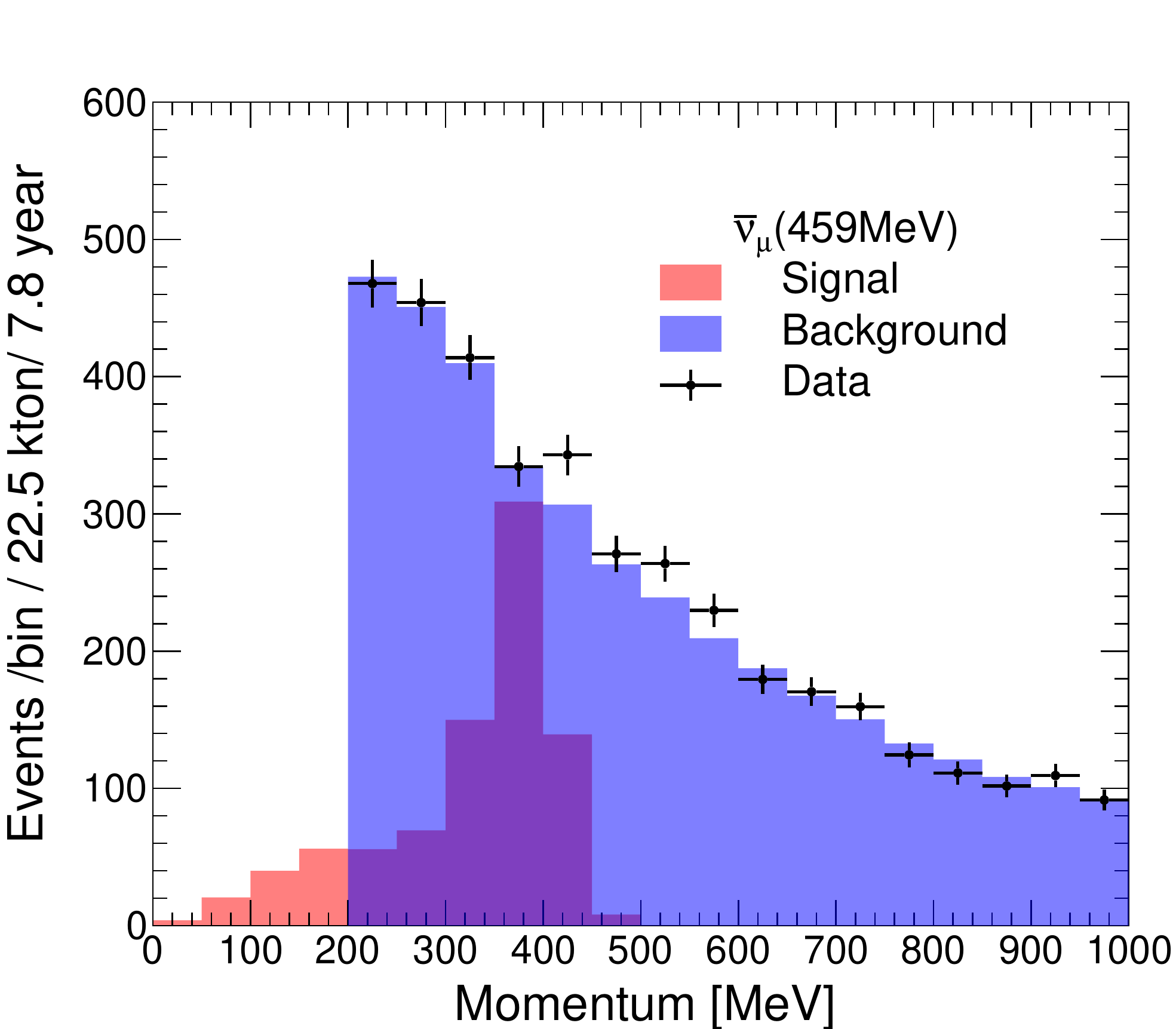}
    \includegraphics[width=0.47\linewidth]{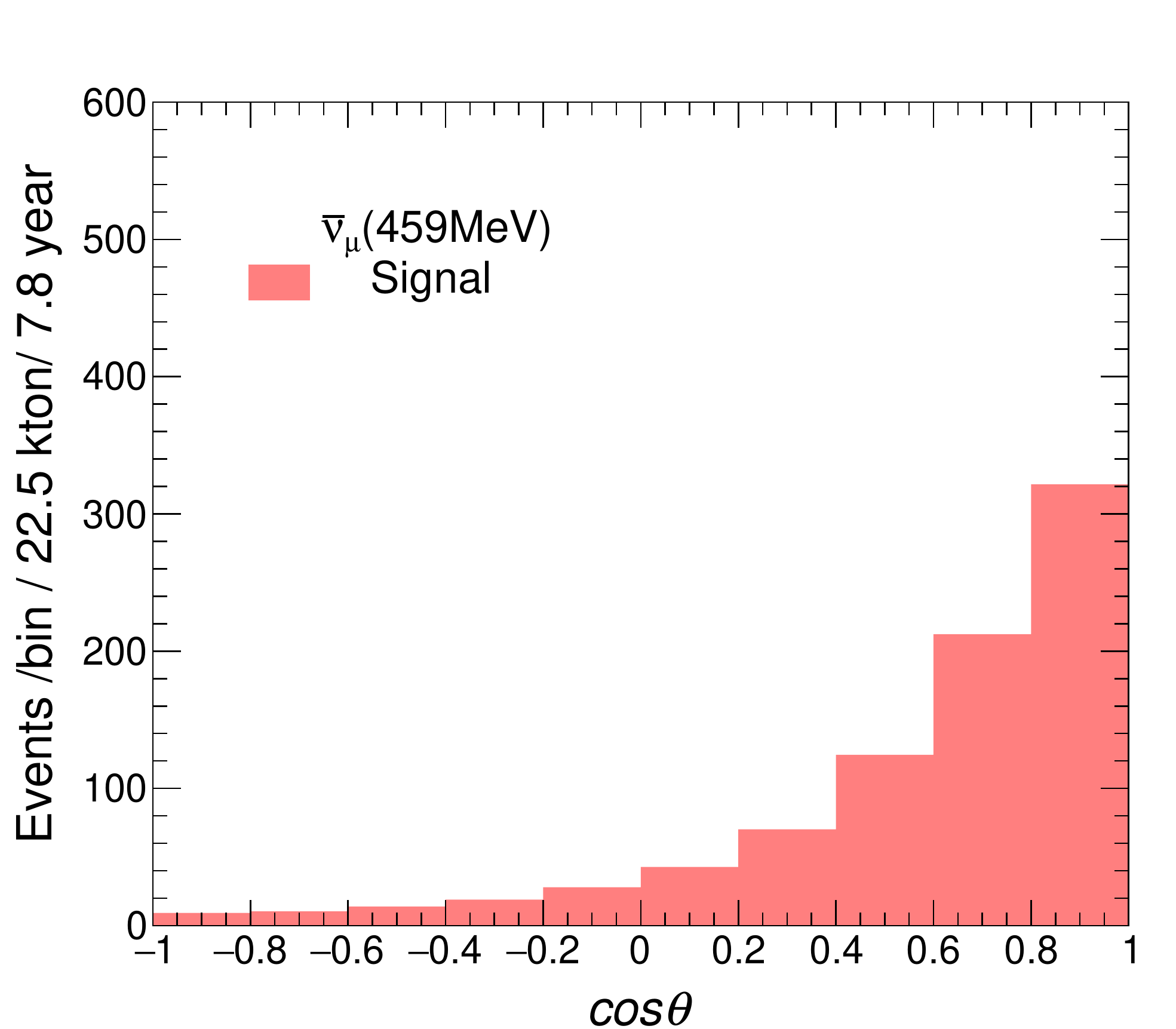}
    \caption{ The expected momentum (left) and angular (right) distributions of charged leptons from the 459 MeV $\bar{\nu}_e$ (upper panels) and $\bar{\nu}_\mu$ (lower panels) CC interactions assuming $f_p/(4\pi R^2) B(p \rightarrow \bar{\nu}_e \, \pi^+) = 1 \, \rm{cm^{-2} s^{-1}}$. The background and observed momentum distributions come from Ref. \cite{Super-Kamiokande:2014pqx}, and are scaled to the level of 176 kton$\cdot$year exposure.} 
    \label{nu-459}
\end{figure}

For the 459 MeV $\bar{\nu}_e$ from $p \rightarrow \bar{\nu}_e \, \pi^+$, we shall analyze both $\bar{\nu}_e$ and $\bar{\nu}_\mu$ CC events in the Super-K detector due to large neutrino oscillation probabilities $P_{\bar\nu_e-\bar\nu_e(\bar\nu_\mu)}(459\rm{MeV})=0.67(0.23)$. For single-ring $e$-like ($\mu$-like) events in Super-K, the momentum and angular resolutions are estimated to be $0.6+2.6/\sqrt{p_l ({\rm GeV})}\%$ ($1.7+0.7/\sqrt{p_l ({\rm GeV})}\%$) and $3.0^\circ$ ($1.8^\circ$) \cite{Super-Kamiokande:2005mbp}, respectively. Based on the GENIE simulations of the 459 MeV $\bar{\nu}_e$ and $\bar{\nu}_\mu$ CC interactions, one can determine the momentum and angular distributions of the final state particles $e^+$ and $\mu^+$ as shown in  Fig. \ref{nu-459}. They have been normalized to the values in Eq. (\ref{Nsignal}). $\theta$ is defined as the angle between the initial neutrino direction (the Sun direction) and the charged lepton direction. Unlike the 236 MeV $\nu_e$ CC interaction, the 459 MeV $\bar{\nu}_e$ and $\bar{\nu}_\mu$ CC events show the directional feature. This is because that antineutrinos will statistically transfer more momentum to the charged lepton than neutrinos in the CC interactions. In  Fig. \ref{nu-459}, the momentum distributions of atmospheric neutrino backgrounds and observed data in Super-K come from Ref. \cite{Super-Kamiokande:2014pqx}, and are scaled to the level of 176 kton$\cdot$year exposure. Assuming they have a uniform angular distribution, we scan the parameter space of the charged lepton momentum $p_l$ and direction $\cos{\theta}$ to maximize the Super-K discovery potential. The optimal selection criteria on $p_l$ and $\cos{\theta}$, and the corresponding signal efficiency $\varepsilon$, background number $N_{\rm bkg}$ and observed number $N_{\rm obs}$, have been listed in Table \ref{summary}.

\begin{table}[]
\setlength{\belowcaptionskip}{0.2cm}
\centering
\caption{ Summary of the analyzed neutrinos, the corresponding selection criteria on momentum $p_l$ and angle $\cos{\theta}$, background number $N_{\rm bkg}$, observed number $N_{\rm obs}$, signal efficiency $\varepsilon$, and $90\%$ C.L. upper limit $N_{90}$ on the expected signal number in Super-K for three proton decay modes. }
\begin{tabular}{c|c|c|c|c|c|c|ccccc} 
\hline \hline
 Decay mode  &  Neutrino    &$p_l \,{\rm (MeV)}$ &$\cos{\theta} $    & $N_{\rm bkg}$ & $N_{\rm obs}$ & $\varepsilon$   & $N_{90}$     \\ \hline
 
 $p \rightarrow e^+ \widetilde\pi \rightarrow \pi^+ $  & $\bar\nu_e (\leq 53)$ & [20, 55]   & -           & 300    & 317    & 0.82    & 42.85  \\ \hline
 
 $p \rightarrow \mu^+ K^0 $   &    $\nu_e$(236)        & [100, 200]   & -           & 1040.6 &1010.1  & 0.87    & 38.58   \\ \hline
 
 \multirow{2}*{$p \rightarrow \bar{\nu}_e \pi^+ $ }  
  &   $\bar\nu_e$(459)    & [350, 450]   & $\geq$ 0.6  & 151.1    & 159.3  & 0.33    &\multirow{2}*{32.65}    \\
    \cline{2-7}
   & $\bar\nu_\mu$(459)  & [350, 400]   & $\geq$ 0.4  & 100.3  &100.4   & 0.28    &    \\ \hline \hline
      \end{tabular}
    \label{summary}
\end{table}

\section{\label{sec:5} The Super-K sensitivities}

To estimate the Super-K sensitivities to three typical proton decay modes, we firstly calculate the $90\%$ confidence level (C.L.) upper limit $N_{90}$ on the expected signal number $N_{\rm s}$ through the following formulas \cite{Super-Kamiokande:2011wjy,Guo:2015hsy}
\begin{equation}
    90\% = \frac{\int_{N_{\rm s}=0}^{N_{90}} L(N_{\rm obs} | N_{\rm s}) dN_{\rm s}}{\int_{N_{\rm s}=0}^{\infty} L(N_{\rm obs} | N_{\rm s}) dN_{\rm s}},
\end{equation}
with the Poisson-based likelihood function

\begin{equation}
    L(N_{\rm obs} | N_{\rm s}) = \prod_{i=1}^{2} \frac{(N_{\rm s} F^i+N_{\rm bkg}^i)^{N_{\rm obs}^i}}{N_{\rm obs}^i!}e^{-(N_{\rm s} F^i+N_{\rm bkg}^i)},
\end{equation}
where the index $i$ refers to the classification of expected signals. $F^i$ denotes the fraction of each category, and can be determined from $N_{\nu_\alpha}^i$ in Eq. (\ref{Nsignal}) and the corresponding $\varepsilon^i$ in Table \ref{summary}. For $p \rightarrow \bar{\nu}_e \pi^+$, two fractions of $F^1 = 0.79$ and $F^2 = 0.21$ refer to the 459 MeV $\bar{\nu}_e$ and $\bar{\nu}_\mu$ categories, respectively. For the other two decay modes, we only analyze a type of signal and take $F^1=1$. With the help of the background number $N_{B}^i$ and observed number $N_{\rm obs}^i$, the corresponding $N_{90}$ of each decay mode has been calculated and listed in the last column of Table \ref{summary}.

\begin{figure}
    \centering
    \includegraphics[width=0.48\linewidth]{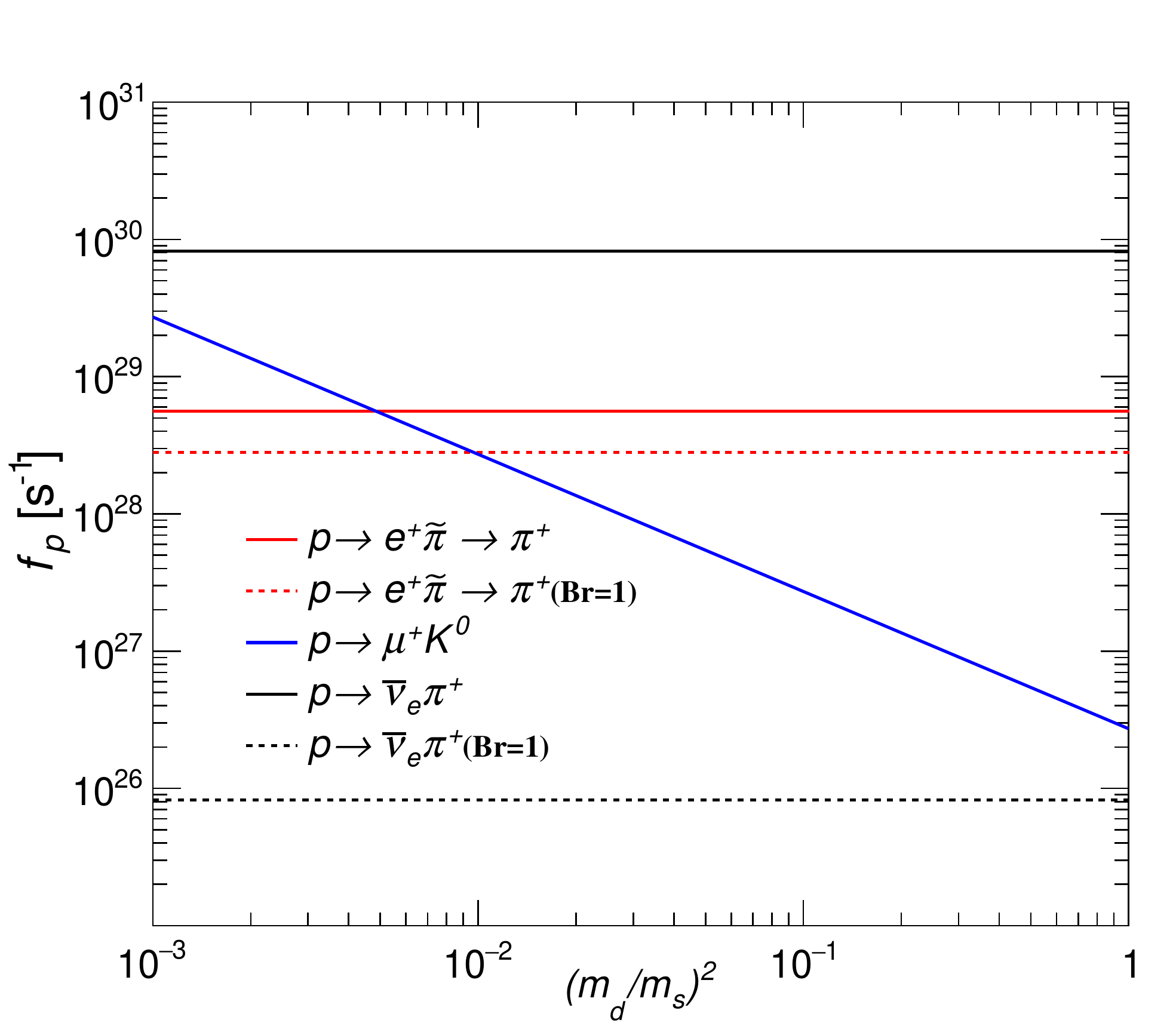}
    \includegraphics[width=0.48\linewidth]{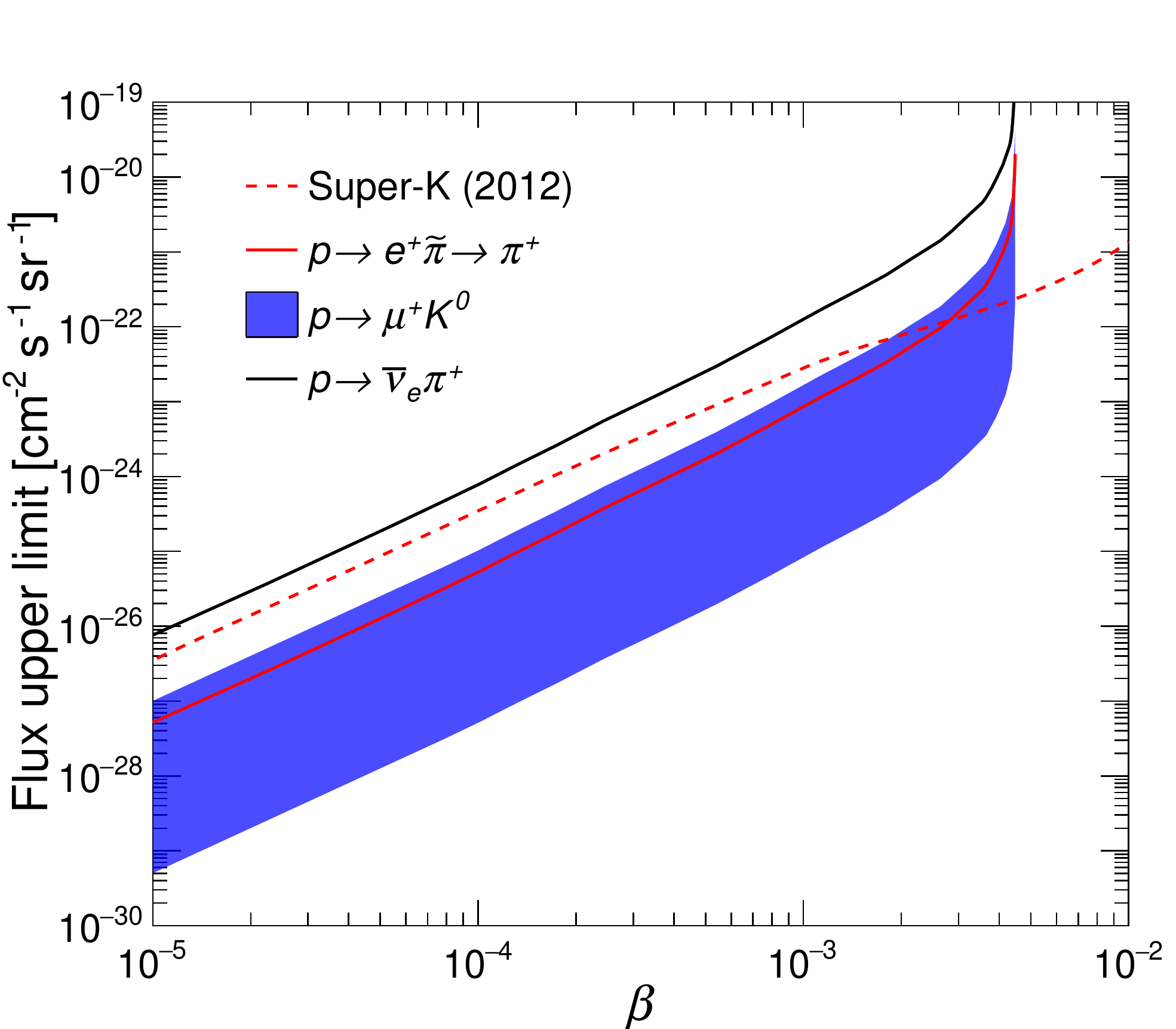}
    \caption{$90$\% C.L. upper limits to the monopole-catalyzed proton decay rate $f_p$ (left panel) in the Sun and the monopole flux (right panel) for the Super-K 176 kton$\cdot$year exposure. Here we take the branching ratios of $B(p \rightarrow e^+ \widetilde\pi \rightarrow \pi^+) = 0.5$, $B(p \rightarrow \mu^+ K^0) \approx {m_d^2}/{m_s^2}$ and $B(p \rightarrow \bar{\nu}_e \, \pi^+)= 10^{-4}$. In the right figure, the catalysis cross section $\sigma_0 =1$ mb, $M_M =10^{16}$ GeV and the Dirac magnetic charge have been assumed for the GUT monopole, the bule band corresponds to ${m_d^2}/{m_s^2}$ in the range of [1/400 - 1/2].}  
    \label{limits}
\end{figure}

Then we use the formula 
\begin{equation}
     N_{90} = \sum_{i=1}^2 N_{\nu_\alpha}^i \, \varepsilon^i 
\label{fplimit}
\end{equation} 
to estimate the $90\%$ C.L. upper limit to the monopole-catalyzed proton decay rate $f_p$ in the Sun. In the left panel of Fig. \ref{limits}, we plot the Super-K upper limits on $f_p$ for three typical proton decay modes, where the branching ratios $B(p \rightarrow e^+ \widetilde\pi \rightarrow \pi^+) = 0.5$, $B(p \rightarrow \mu^+ K^0) = (m_d/m_s)^2 \sim 1/400-1/2$ and $B(p \rightarrow \bar{\nu}_e \pi^+) = 10^{-4}$ have been used. For reference, two dashed lines corresponding to the $B(p \rightarrow e^+ \widetilde\pi \rightarrow \pi^+) = 1$ and $B(p \rightarrow \bar{\nu}_e \, \pi^+)= 1$ cases have also been added. If we consider the contribution from the momentum [100,350] ([200,350]) MeV bins for the 459 MeV $\bar{\nu}_e$ ($\bar{\nu}_\mu$) analysis, the limits will be improved by $4\%$ ($10\%$). For 236 MeV $\nu_e$ neutrinos, the improvement is not apparent when the [200,250] MeV bin is included. It is clear that the decay mode $p \rightarrow \mu^+ K^0$ always give a better limit than $p \rightarrow \bar{\nu}_e \pi^+$. This is because that $p \rightarrow \mu^+ K^0$  has the larger $B \, Y_{\nu_l} = [1/400-1/2] \times 0.21 \approx [1/2000-1/10]$ in Eq. (\ref{flux}) than $10^{-4}$ from the $p \rightarrow \bar{\nu}_e \pi^+$ mode for the high energy neutrino production. In addition, the momentum distributions of charged leptons from the 459 MeV antineutrino CC interactions have the larger smearing than the 236 MeV neutrino as shown in Figs. \ref{nue-236} and \ref{nu-459}. For $B(p \rightarrow \mu^+ K^0) = (m_d/m_s)^2 > 5.0 \times 10^{-3}$, the best experimental limit will come from $p \rightarrow \mu^+ K^0$ among all three proton decay modes. Compared with the $p \rightarrow e^+ \widetilde\pi \rightarrow \pi^+$ mode, the $p \rightarrow \mu^+ K^0$ mode can produce the 236 MeV $\nu_\mu$ neutrino and can also give the 236 MeV $\nu_e$ neutrino through the neutrino oscillation. It is found that the 236 MeV $\nu_e$ neutrino has the larger CC cross section $2.51 \times 10^{-38} {\rm cm^2}$ \cite{Andreopoulos:2009rq} than the IBD cross section $\sim 10^{-41}-10^{-40} {\rm cm^2}$ \cite{Strumia:2003zx} for the low energy $\bar{\nu}_e$ antineutrino. So we suggest the Super-K collaboration searches for this important proton decay mode in the future analysis. Assuming the same background level and distribution, we calculate the expected Super-K sensitivity with the full available 359 kton$\cdot$year data-set \cite{Super-Kamiokande:2021jaq}. Compared with the limits in Fig. \ref{limits} from the 176 kton$\cdot$year exposure, the expected sensitives of $p \rightarrow e^+ \widetilde\pi \rightarrow \pi^+$, $p \rightarrow \mu^+ K^0$, $p \rightarrow \bar{\nu}_e \pi^+$ from the 359 kton$\cdot$year exposure will be improved by $21\%$, $40\%$ and $24\%$, respectively. Note that the Hyper-Kamiokande experiment \cite{Abe:2011ts} has the ability to give a better limit due to its huge target mass.

Finally, we calculate the Super-K upper limit on the monopole flux by use of Eqs. (\ref{eq2}) and (\ref{eq4}) as shown in the right panel of Fig. \ref{limits}. The predicted monopole flux limits will become weaker as the monopole velocity $\beta$ increases. For comparison, we also show the Super-K published limit (dashed line) \cite{Super-Kamiokande:2012tld} in this figure. Note that it has been 
divided by a factor of 0.17 because of $\sigma_R = \sigma_0/ \beta^2_{\rm rel}$ used in Ref. \cite{Super-Kamiokande:2012tld} to describe the cross section of monopole-catalyzed proton decays, rather than $\sigma_R = 0.17 \sigma_0/ \beta^2_{\rm rel}$ used in this paper. We find that the predicted and published limits have different behaviors for $\beta > 4.5 \times 10^{-3}$, which is caused by different results used for the capture fraction $\epsilon (M_M,\beta,g)$ in Eq. (\ref{eq2}). As discussed in Sec. \ref{sec:2}, $\epsilon (M_M,\beta,g)$ depends on the energy loss rate of monopoles passing through the Sun. The $\epsilon (M_M,\beta,g)$ calculation in Ref. \cite{Super-Kamiokande:2012tld} is based on the collective effects which can enhance the monopole energy loss rate \cite{Hamilton:1983qme}. Therefore, the Super-K has a limit on the monopole flux for $\beta > 4.5 \times 10^{-3}$ since these monopoles can be captured by the Sun due to the large stopping power. However, other calculations find that the energy loss from collective effects is insignificant \cite{Meyer-Vernet:1985yyn,Bracci:1984zr}, and about one order smaller than that used in this paper \cite{Ahlen:1996ax}. So we do not consider the contribution from collective effects to the monopole stopping power. On the other hand, the predicted result (red line) is stronger than the published one (dashed line) if $\beta < 3 \times 10^{-3}$ for the proton decay mode $p \rightarrow e^+ \widetilde\pi \rightarrow \pi^+$. This is because that we ignore the relevant uncertainties and use a different method to roughly calculate the sensitivity. Note that this difference does not change the conclusion that the proton decay mode $p \rightarrow \mu^+ K^0$ can give the best limit among the three decay modes for most of the parameter space.

\section{\label{sec:6} Conclusions }

In conclusion, we have investigated the neutrino signals from proton decays catalyzed by GUT monopoles in the Sun. Three typical proton decay modes, $p \rightarrow e^+ \widetilde\pi$, $p \rightarrow \mu^+ K^0$ and $p \rightarrow \bar{\nu}_e \pi^+$,  have been analyzed for the Super-K experiment. To obtain the neutrino energy spectra, we use the Geant4 software to simulate interactions of decay products in the highly-dense solar center. It is found that $K^0$ can produce a large amount of 236 MeV monoenergetic $\nu_\mu$ neutrinos through the charge exchange process $K^0 + p \rightarrow K^+ + n$ and the subsequent decay $K^+ \rightarrow \mu^+ \nu_\mu$. This interesting feature is not realized in the previous papers. Based on the signal and background distributions of three proton decay modes, we set the reasonable selection conditions, and estimate the corresponding signal efficiencies and backgrounds in Super-K. Then we calculate the Super-K upper limit on the monopole-catalyzed proton decay rate $f_p$, and find $p \rightarrow \mu^+ K^0$ can give the best limit among three decay modes for most of the parameter space. Note that the decay mode $p \rightarrow \mu^+ K^0$ always give a better limit than $p \rightarrow \bar{\nu}_e \pi^+$. So we suggest the Super-K collaboration searches for this important proton decay mode in the future analysis. Finally, we present the Super-K sensitivities to the monopole flux for three proton decay modes.

\acknowledgments

This work is supported in part by the National Nature Science Foundation of China (NSFC) under Grants No. 11575201 and No. 11675273, and the Strategic Priority Research Program of the Chinese Academy of Sciences under Grant No. XDA10010100.

%%%%%% Reference


\begin{thebibliography}{99}

%\cite{tHooft:1974kcl}
\bibitem{tHooft:1974kcl}
G.~'t Hooft,
%``Magnetic Monopoles in Unified Gauge Theories,''
Nucl. Phys. B \textbf{79} (1974), 276-284
doi:10.1016/0550-3213(74)90486-6
%3200 citations counted in INSPIRE as of 19 Dec 2021


%\cite{Polyakov:1974ek}
\bibitem{Polyakov:1974ek}
A.~M.~Polyakov,
%``Particle Spectrum in Quantum Field Theory,''
JETP Lett. \textbf{20} (1974), 194-195
PRINT-74-1566 (LANDAU-INST).
%2509 citations counted in INSPIRE as of 19 Dec 2021


%\cite{ParticleDataGroup:2020ssz}
\bibitem{ParticleDataGroup:2020ssz}
P.~A.~Zyla \textit{et al.} [Particle Data Group],
%``Review of Particle Physics,''
PTEP \textbf{2020} (2020) no.8, 083C01
doi:10.1093/ptep/ptaa104
%2491 citations counted in INSPIRE as of 19 Dec 2021


%\cite{Burdin:2014xma}
\bibitem{Burdin:2014xma}
S.~Burdin, M.~Fairbairn, P.~Mermod, D.~Milstead, J.~Pinfold, T.~Sloan and W.~Taylor,
%``Non-collider searches for stable massive particles,''
Phys. Rept. \textbf{582} (2015), 1-52
doi:10.1016/j.physrep.2015.03.004
[arXiv:1410.1374 [hep-ph]].
%78 citations counted in INSPIRE as of 19 Dec 2021


%\cite{Patrizii:2015uea}
\bibitem{Patrizii:2015uea}
L.~Patrizii and M.~Spurio,
%``Status of Searches for Magnetic Monopoles,''
Ann. Rev. Nucl. Part. Sci. \textbf{65} (2015), 279-302
doi:10.1146/annurev-nucl-102014-022137
[arXiv:1510.07125 [hep-ex]].
%50 citations counted in INSPIRE as of 19 Dec 2021


%\cite{Mavromatos:2020gwk}
\bibitem{Mavromatos:2020gwk}
N.~E.~Mavromatos and V.~A.~Mitsou,
%``Magnetic monopoles revisited: Models and searches at colliders and in the Cosmos,''
Int. J. Mod. Phys. A \textbf{35} (2020) no.23, 2030012
doi:10.1142/S0217751X20300124
[arXiv:2005.05100 [hep-ph]].
%25 citations counted in INSPIRE as of 19 Dec 2021


%\cite{Rubakov:1982fp}
\bibitem{Rubakov:1982fp}
V.~A.~Rubakov,
%``Adler-Bell-Jackiw Anomaly and Fermion Number Breaking in the Presence of a Magnetic Monopole,''
Nucl. Phys. B \textbf{203} (1982), 311-348
doi:10.1016/0550-3213(82)90034-7
%528 citations counted in INSPIRE as of 19 Dec 2021


%\cite{Callan:1982au}
\bibitem{Callan:1982au}
C.~G.~Callan, Jr.,
%``Dyon-Fermion Dynamics,''
Phys. Rev. D \textbf{26} (1982), 2058-2068
doi:10.1103/PhysRevD.26.2058
%489 citations counted in INSPIRE as of 19 Dec 2021


%\cite{Kajita:1985aig}
\bibitem{Kajita:1985aig}
T.~Kajita, K.~Arisaka, M.~Koshiba, M.~Nakahata, Y.~Oyama, A.~Suzuki, M.~Takita, Y.~Totsuka, T.~Kifune and T.~Suda, \textit{et al.}
%``SEARCH FOR NUCLEON DECAYS CATALYZED BY MAGNETIC MONOPOLES,''
J. Phys. Soc. Jap. \textbf{54} (1985), 4065-4068
doi:10.1143/JPSJ.54.4065
%33 citations counted in INSPIRE as of 19 Dec 2021


%\cite{Bartelt:1986cv}
\bibitem{Bartelt:1986cv}
J.~E.~Bartelt, H.~Courant, K.~J.~Heller, T.~Joyce, M.~Marshak, E.~Peterson, K.~Ruddick, M.~Shupe, D.~S.~Ayres and J.~W.~Dawson, \textit{et al.}
%``Monopole Flux and Proton Decay Limits From the Soudan-i Detector,''
Phys. Rev. D \textbf{36} (1987), 1990
[erratum: Phys. Rev. D \textbf{40} (1989), 1701]
doi:10.1103/PhysRevD.36.1990
%36 citations counted in INSPIRE as of 19 Dec 2021


%\cite{Becker-Szendy:1994kqw}
\bibitem{Becker-Szendy:1994kqw}
R.~Becker-Szendy, C.~B.~Bratton, J.~Breault, D.~Casper, S.~T.~Dye, K.~Ganezer, W.~Gajewski, M.~Goldhaber, T.~J.~Haines and P.~G.~Halverson, \textit{et al.}
%``New magnetic monopole flux limits from the IMB proton decay detector,''
Phys. Rev. D \textbf{49} (1994), 2169-2173
doi:10.1103/PhysRevD.49.2169
%22 citations counted in INSPIRE as of 19 Dec 2021


%\cite{Baikal:1997kuo}
\bibitem{Baikal:1997kuo}
V.~A.~Balkanov \textit{et al.} [Baikal],
%``The Baikal deep underwater neutrino experiment: Results, status, future,''
Prog. Part. Nucl. Phys. \textbf{40} (1998), 391-401
doi:10.1016/S0146-6410(98)00047-7
[arXiv:astro-ph/9801044 [astro-ph]].
%42 citations counted in INSPIRE as of 19 Dec 2021


%\cite{MACRO:2002iaq}
\bibitem{MACRO:2002iaq}
M.~Ambrosio \textit{et al.} [MACRO],
%``Search for nucleon decays induced by GUT magnetic monopoles with the MACRO experiment,''
Eur. Phys. J. C \textbf{26} (2002), 163-172
doi:10.1140/epjc/s2002-01045-x
[arXiv:hep-ex/0207024 [hep-ex]].
%65 citations counted in INSPIRE as of 19 Dec 2021


%\cite{IceCube:2014xnp}
\bibitem{IceCube:2014xnp}
M.~G.~Aartsen \textit{et al.} [IceCube],
%``Search for non-relativistic Magnetic Monopoles with IceCube,''
Eur. Phys. J. C \textbf{74} (2014) no.7, 2938
[erratum: Eur. Phys. J. C \textbf{79} (2019) no.2, 124]
doi:10.1140/epjc/s10052-014-2938-8
[arXiv:1402.3460 [astro-ph.CO]].
%56 citations counted in INSPIRE as of 19 Dec 2021

%\cite{Kolb:1982si}
\bibitem{Kolb:1982si}
E.~W.~Kolb, S.~A.~Colgate and J.~A.~Harvey,
%``Monopole Catalysis of Nucleon Decay in Neutron Stars,''
Phys. Rev. Lett. \textbf{49} (1982), 1373
doi:10.1103/PhysRevLett.49.1373
%141 citations counted in INSPIRE as of 19 Dec 2021


%\cite{Freese:1998es}
\bibitem{Freese:1998es}
K.~Freese and E.~Krasteva,
%``A Bound on the flux of magnetic monopoles from catalysis of nucleon decay in white dwarfs,''
Phys. Rev. D \textbf{59} (1999), 063007
doi:10.1103/PhysRevD.59.063007
[arXiv:astro-ph/9804148 [astro-ph]].
%12 citations counted in INSPIRE as of 19 Dec 2021


%\cite{Arafune:1983tr}
\bibitem{Arafune:1983tr}
J.~Arafune, M.~Fukugita and S.~Yanagita,
%``Monopole Abundance in the Solar System and the Intrinsic Heat in the Jovian Planets,''
Phys. Rev. D \textbf{32} (1985), 2586
doi:10.1103/PhysRevD.32.2586
%13 citations counted in INSPIRE as of 19 Dec 2021

%\cite{Super-Kamiokande:2012tld}
\bibitem{Super-Kamiokande:2012tld}
K.~Ueno \textit{et al.} [Super-Kamiokande],
%``Search for GUT monopoles at Super\textendash{}Kamiokande,''
Astropart. Phys. \textbf{36} (2012), 131-136
doi:10.1016/j.astropartphys.2012.05.008
[arXiv:1203.0940 [hep-ex]].
%34 citations counted in INSPIRE as of 19 Dec 2021


%\cite{Bais:1982hm}
\bibitem{Bais:1982hm}
F.~A.~Bais, J.~R.~Ellis, D.~V.~Nanopoulos and K.~A.~Olive,
%``MORE ABOUT BARYON NUMBER VIOLATION CATALYZED BY GRAND UNIFIED MONOPOLES,''
Nucl. Phys. B \textbf{219} (1983), 189-219
doi:10.1016/0550-3213(83)90434-0
%49 citations counted in INSPIRE as of 19 Dec 2021


%\cite{Houston:2018rvz}
\bibitem{Houston:2018rvz}
N.~Houston, T.~Li and C.~Sun,
%``A new solar neutrino channel for grand-unification monopole searches,''
JCAP \textbf{10} (2018), 034
doi:10.1088/1475-7516/2018/10/034
[arXiv:1803.02835 [hep-ph]].
%1 citations counted in INSPIRE as of 19 Dec 2021

%\cite{Kibble:1976sj}
\bibitem{Kibble:1976sj}
T.~W.~B.~Kibble,
%``Topology of Cosmic Domains and Strings,''
J. Phys. A \textbf{9} (1976), 1387-1398
doi:10.1088/0305-4470/9/8/029
%2653 citations counted in INSPIRE as of 20 Dec 2021

%\cite{Parker:1970xv}
\bibitem{Parker:1970xv}
E.~N.~Parker,
%``The Origin of Magnetic Fields,''
Astrophys. J. \textbf{160} (1970), 383
doi:10.1086/150442
%293 citations counted in INSPIRE as of 19 Dec 2021

%\cite{Turner:1982ag}
\bibitem{Turner:1982ag}
M.~S.~Turner, E.~N.~Parker and T.~J.~Bogdan,
%``Magnetic Monopoles and the Survival of Galactic Magnetic Fields,''
Phys. Rev. D \textbf{26} (1982), 1296
doi:10.1103/PhysRevD.26.1296
%255 citations counted in INSPIRE as of 20 Dec 2021

%\cite{Frieman:1985dv}
\bibitem{Frieman:1985dv}
J.~A.~Frieman, K.~Freese and M.~S.~Turner,
%``Superheavy Magnetic Monopoles and Main Sequence Stars,''
Astrophys. J. \textbf{335} (1988), 844-861
doi:10.1086/166972
%11 citations counted in INSPIRE as of 19 Dec 2021


%\cite{Ahlen:1996ax}
\bibitem{Ahlen:1996ax}
S.~P.~Ahlen, I.~De Mitri, J.~T.~Hong and G.~Tarle,
%``Energy loss of supermassive magnetic monopoles and dyons in main sequence stars,''
Phys. Rev. D \textbf{55} (1997), 6584-6590
doi:10.1103/PhysRevD.55.6584
%4 citations counted in INSPIRE as of 19 Dec 2021


%\cite{Ahlen:1996ay}
\bibitem{Ahlen:1996ay}
S.~P.~Ahlen, I.~De Mitri, J.~T.~Hong and G.~Tarle,
%``Superheavy magnetic monopoles trapped into the sun and the solar system,''
INFN-AE-96-34.
%0 citations counted in INSPIRE as of 19 Dec 2021


%\cite{Arafune:1983uz}
\bibitem{Arafune:1983uz}
J.~Arafune and M.~Fukugita,
%``Velocity Dependent Factors for the Rubakov Process for Slowly Moving Magnetic Monopoles in Matter,''
Phys. Rev. Lett. \textbf{50} (1983), 1901
doi:10.1103/PhysRevLett.50.1901
%57 citations counted in INSPIRE as of 19 Dec 2021

%\cite{Vinyoles:2016djt}
\bibitem{Vinyoles:2016djt}
N.~Vinyoles, A.~M.~Serenelli, F.~L.~Villante, S.~Basu, J.~Bergstr\"om, M.~C.~Gonzalez-Garcia, M.~Maltoni, C.~Pe\~na-Garay and N.~Song,
%``A new Generation of Standard Solar Models,''
Astrophys. J. \textbf{835} (2017) no.2, 202
doi:10.3847/1538-4357/835/2/202
[arXiv:1611.09867 [astro-ph.SR]].
%180 citations counted in INSPIRE as of 19 Dec 2021

%\cite{Arafune:1983sk}
\bibitem{Arafune:1983sk}
J.~Arafune and M.~Fukugita,
%``Limit on the Solar Monopole Abundance,''
Phys. Lett. B \textbf{133} (1983), 380-384
doi:10.1016/0370-2693(83)90810-9
%20 citations counted in INSPIRE as of 19 Dec 2021


%\cite{Baratella:2013fya}
\bibitem{Baratella:2013fya}
P.~Baratella, M.~Cirelli, A.~Hektor, J.~Pata, M.~Piibeleht and A.~Strumia,
%``PPPC 4 DM$\nu$: a Poor Particle Physicist Cookbook for Neutrinos from Dark Matter annihilations in the Sun,''
JCAP \textbf{03} (2014), 053
doi:10.1088/1475-7516/2014/03/053
[arXiv:1312.6408 [hep-ph]].
%72 citations counted in INSPIRE as of 19 Dec 2021


%\cite{GEANT4:2002zbu}
\bibitem{GEANT4:2002zbu}
S.~Agostinelli \textit{et al.} [GEANT4],
%``GEANT4--a simulation toolkit,''
Nucl. Instrum. Meth. A \textbf{506} (2003), 250-303
doi:10.1016/S0168-9002(03)01368-8
%14710 citations counted in INSPIRE as of 19 Dec 2021

%\cite{Wright:2015xia}
\bibitem{Wright:2015xia}
D.~H.~Wright and M.~H.~Kelsey,
%``The Geant4 Bertini Cascade,''
Nucl. Instrum. Meth. A \textbf{804} (2015), 175-188
doi:10.1016/j.nima.2015.09.058
%50 citations counted in INSPIRE as of 09 Apr 2022

%\cite{Bohlen:2014buj}
\bibitem{Bohlen:2014buj}
T.~T.~B\"ohlen, F.~Cerutti, M.~P.~W.~Chin, A.~Fass\`o, A.~Ferrari, P.~G.~Ortega, A.~Mairani, P.~R.~Sala, G.~Smirnov and V.~Vlachoudis,
%``The FLUKA Code: Developments and Challenges for High Energy and Medical Applications,''
Nucl. Data Sheets \textbf{120} (2014), 211-214
doi:10.1016/j.nds.2014.07.049
%665 citations counted in INSPIRE as of 21 Dec 2021


%\cite{Guo:2018sno}
\bibitem{Guo:2018sno}
W.~L.~Guo,
%``Low energy neutrinos from stopped muons in the Earth,''
Phys. Rev. D \textbf{99} (2019) no.7, 073007
doi:10.1103/PhysRevD.99.073007
[arXiv:1812.04378 [hep-ph]].
%6 citations counted in INSPIRE as of 19 Dec 2021


%\cite{Strumia:2003zx}
\bibitem{Strumia:2003zx}
A.~Strumia and F.~Vissani,
%``Precise quasielastic neutrino/nucleon cross-section,''
Phys. Lett. B \textbf{564} (2003), 42-54
doi:10.1016/S0370-2693(03)00616-6
[arXiv:astro-ph/0302055 [astro-ph]].
%380 citations counted in INSPIRE as of 19 Dec 2021


%\cite{Andreopoulos:2009rq}
\bibitem{Andreopoulos:2009rq}
C.~Andreopoulos, A.~Bell, D.~Bhattacharya, F.~Cavanna, J.~Dobson, S.~Dytman, H.~Gallagher, P.~Guzowski, R.~Hatcher and P.~Kehayias, \textit{et al.}
%``The GENIE Neutrino Monte Carlo Generator,''
Nucl. Instrum. Meth. A \textbf{614} (2010), 87-104
doi:10.1016/j.nima.2009.12.009
[arXiv:0905.2517 [hep-ph]].
%918 citations counted in INSPIRE as of 19 Dec 2021


%\cite{Super-Kamiokande:2005mbp}
\bibitem{Super-Kamiokande:2005mbp}
Y.~Ashie \textit{et al.} [Super-Kamiokande],
%``A Measurement of atmospheric neutrino oscillation parameters by SUPER-KAMIOKANDE I,''
Phys. Rev. D \textbf{71} (2005), 112005
doi:10.1103/PhysRevD.71.112005
[arXiv:hep-ex/0501064 [hep-ex]].
%1194 citations counted in INSPIRE as of 23 Dec 2021

%\cite{Super-Kamiokande:2014pqx}
\bibitem{Super-Kamiokande:2014pqx}
V.~Takhistov \textit{et al.} [Super-Kamiokande],
%``Search for Trilepton Nucleon Decay via $p \rightarrow e^+ \nu \nu$ and $p \rightarrow \mu^+ \nu \nu$ in the Super-Kamiokande Experiment,''
Phys. Rev. Lett. \textbf{113} (2014) no.10, 101801
doi:10.1103/PhysRevLett.113.101801
[arXiv:1409.1947 [hep-ex]].
%29 citations counted in INSPIRE as of 19 Dec 2021

%\cite{Super-Kamiokande:2011wjy}
\bibitem{Super-Kamiokande:2011wjy}
T.~Tanaka \textit{et al.} [Super-Kamiokande],
%``An Indirect Search for WIMPs in the Sun using 3109.6 days of upward-going muons in Super-Kamiokande,''
Astrophys. J. \textbf{742} (2011), 78
doi:10.1088/0004-637X/742/2/78
[arXiv:1108.3384 [astro-ph.HE]].
%214 citations counted in INSPIRE as of 27 Dec 2021

%\cite{Guo:2015hsy}
\bibitem{Guo:2015hsy}
W.~L.~Guo,
%``Detecting electron neutrinos from solar dark matter annihilation by JUNO,''
JCAP \textbf{01} (2016), 039
doi:10.1088/1475-7516/2016/01/039
[arXiv:1511.04888 [hep-ph]].
%7 citations counted in INSPIRE as of 27 Dec 2021

%\cite{Super-Kamiokande:2021jaq}
\bibitem{Super-Kamiokande:2021jaq}
K.~Abe \textit{et al.} [Super-Kamiokande],
%``Diffuse supernova neutrino background search at Super-Kamiokande,''
Phys. Rev. D \textbf{104} (2021) no.12, 122002
doi:10.1103/PhysRevD.104.122002
[arXiv:2109.11174 [astro-ph.HE]].
%17 citations counted in INSPIRE as of 05 May 2022

%\cite{Abe:2011ts}
\bibitem{Abe:2011ts}
K.~Abe, T.~Abe, H.~Aihara, Y.~Fukuda, Y.~Hayato, K.~Huang, A.~K.~Ichikawa, M.~Ikeda, K.~Inoue and H.~Ishino, \textit{et al.}
%``Letter of Intent: The Hyper-Kamiokande Experiment --- Detector Design and Physics Potential ---,''
[arXiv:1109.3262 [hep-ex]].
%666 citations counted in INSPIRE as of 10 Apr 2022

%\cite{Hamilton:1983qme}
\bibitem{Hamilton:1983qme}
A.~J.~S.~Hamilton and C.~L.~Sarazin,
%``DECELERATION OF GRAND UNIFIED THEORY MONOPOLES IN A PLASMA,''
Astrophys. J. \textbf{274} (1983), 399-407
doi:10.1086/161455
%9 citations counted in INSPIRE as of 26 Dec 2021

%\cite{Meyer-Vernet:1985yyn}
\bibitem{Meyer-Vernet:1985yyn}
N.~Meyer-Vernet,
%``Energy loss by slow magnetic monopoles in a thermal plasma,''
Astrophys. J. \textbf{290} (1985), 21-23
doi:10.1086/162954
%11 citations counted in INSPIRE as of 26 Dec 2021

%\cite{Bracci:1984zr}
\bibitem{Bracci:1984zr}
L.~Bracci, G.~Fiorentini and G.~Mezzorani,
%``Monopole Trapping Inside Stars and Phenomenological Consequences,''
Nucl. Phys. B \textbf{258} (1985), 726-746
doi:10.1016/0550-3213(85)90633-9
%8 citations counted in INSPIRE as of 26 Dec 2021

\end{thebibliography}
\end{document}